\documentclass[aps,prb,twocolumn,floats]{revtex4}

\usepackage{amssymb}
\usepackage{amsmath}
\usepackage{graphicx}
\usepackage{bm}
\usepackage{mdframed}
\usepackage{color}
\usepackage{gensymb}

\begin{document}

\bibliographystyle{prsty}

\title{Stability and dynamics of in-plane skyrmions in collinear ferromagnets}

\author{Ricardo Zarzuela,$^{1}$ Venkata Krishna Bharadwaj,$^{1}$  Kyoung-Whan Kim,$^{2}$ Jairo Sinova,$^{1,3}$ and Karin Everschor-Sitte$^{1}$}

\affiliation{$^{1}$ Institut f\"{u}r Physik, Johannes Gutenberg Universit\"{a}t Mainz, D-55099 Mainz, Germany \\
$^{2}$ Center for Spintronics, Korea Institute of Science and Technology, Seoul 02792, Korea\\
$^{3}$ Institute of Physics Academy of Sciences of the Czech Republic, Cukrovarnick\'{a} 10, 162 00 Praha 6, Czech Republic}

\begin{abstract}
We study the emergence and dynamics of in-plane skyrmions in collinear ferromagnetic heterostructures. We present a minimal energy model for this class of magnetic textures, determine the crystal symmetries compatible with it and propose material candidates, based on symmetries only, for the observation of these topological solitons. We calculate exact solutions of the energy model for in-plane skyrmions in the absence of dipolar interactions at critical coupling, the latter defined by the relations $H=K$ and $D=\sqrt{AK}$ for the strength of the external magnetic field and the Dzyaloshinskii coupling constant, respectively, with $K$ and $A$ being the anisotropy constant and the exchange stiffness of the material. Through micromagnetic simulations, we demonstrate the possibility of in-plane skyrmion production via i) the motion of domain walls through a geometrical constriction and ii) shedding from a magnetic impurity driven both by spin transfer torques. In-plane skyrmion dynamics triggered by spin-orbit torques are also investigated analytically and numerically. Our findings point towards the possibility of designing racetracks for in-plane skyrmions, whose speed could be tuned by adjusting the angle between the charge current and the uniform background magnetization; in particular, the speed is maximum for currents parallel to the easy axis and becomes zero for currents transverse to it.

\end{abstract}
\maketitle

\section{Introduction} 

Magnetic skyrmions\cite{Belavin-JETP1975} epitomize the class of spatially localized solitons in two dimensions and arise in magnetic systems with spin-orbit coupling and broken centrosymmetry.\cite{Bogdanov-PRB2002,Banerjee-PRX2014} These topological textures have been observed in both crystal\cite{Mulbauer-Sci2009,Yu-Nat2010,Seki-Sci2012} and gas\cite{Romming-Sci2013,Jiang-Sci2015} phases, and can be stabilized by chiral (Dzyaloshinskii-Moriya) interactions,\cite{Dzyaloshinskii-JETP1957,Moriya-PR1960,Dzyaloshinskii-JETP1964,Fert-PRL1980} stray fields, geometric frustration\cite{Leonov-NatComms2015,Lin-PRB2016} and the Ruderman-Kittel-Kasuya-Yosida interaction.\cite{Bezvershenko-PRB2019} Recent years have witnessed a growing interest in their potential usage as building blocks for information processing/storage\cite{Parkin-Science2008,Nagaosa-NatNano2013,Zhou-NatComm2014,Zhang-SciRep2015,Fert-NatRevMat2017} and novel computation paradigms,\cite{Huang-Nanotech2017,Li-Nanotech2017,Prychynenko-PRAppl2018,Azam-JApplPhys2018,Bourianoff-AIPAdv2018} including the skyrmion reshuffler,\cite{Pinna2019, Zazvorka2019} since skyrmions show a particlelike behavior\cite{Pollath-PRL2017,Lin-PRB2013} and a low threshold for current-driven mobility,\cite{Jonietz-Science2010,Schulz-NatPhys2012,Fert-NatNano2013} can be nucleated/annihilated via spin torques\cite{Jiang-Sci2015,Sitte-PRB2016,Everschor-Sitte-NJP2017,Stier-PRL2017,Buttner-NNano2017} or the local injection of spin-polarized currents,\cite{Sampaio-NatNano2013} are robust against external electromagnetic perturbations and structural distortions (e.g., impurities),\cite{Iwasaki-NatComms2013,Rohart-PRB2016} exhibit unconventional dynamical features such as the skyrmion Hall effect\cite{Jiang-NatPhys2017,Litzius-NatPhys2017} and mediate the topological Hall effect in the conducting systems.\cite{Bruno-PRL2004,Neubauer-PRL2009,Schulz-NatPhys2012}

Experimental platforms usually utilized in current-driven skyrmion transport measurements\cite{} consist of thin-film heterostructures made of a magnet deposited on a (heavy-)metal substrate, the latter endowing an interfacial Dzyaloshinski-Moriya (DM) interaction and a perpendicular (to the basal plane) magnetocrystalline anisotropy (PMA) in the former,\cite{Fert-PRL1980} which promote the stabilization of N\'{e}el skyrmions. In this regard, it is worth noting the existence of other Lifshitz invariants stabilizing skyrmion configurations that can be smoothly deformed into the N\'{e}el-like one by a suitable global rotation in spin space. {\it In-plane skyrmions} (also known as magnetic bimerons), which belong to this broad family, have been recently observed forming a (disordered) lattice in MnSi thin films\cite{Yokouchi-JPSJ2015,Meynell-PRB2017} and are predicted to exist individually in frustrated magnets\cite{Kharkov-PRL2017} in the presence of magnetic fields. These spin textures, which consist of a vortex-antivortex pair, each carrying half the quantum of topological charge,\cite{FN0} are attracting much attention from the spintronics community since they exhibit intrinsically distinct current-driven dynamics to those found for N\'{e}el skyrmions,\cite{Gobel-PRB2019,Shen-2019} and therefore offer alternative conduits for the transport of spin signals and new perspectives for the design of the next generation of high-speed electronic devices.

In this paper we conduct a thorough study of the creation, stability and dynamical properties of in-plane skyrmions. The structure of the manuscript is the following: in Sec. II we present a minimal energy model for the stabilization of in-plane skyrmions, analyze those (crystallographic) space groups compatible with it and propose material candidates for their observation (Sec.~II A). This constitutes the first main finding of our work. We note in passing that, to the best of our knowledge, single in-plane skyrmions can only exist as transient states in magnetic systems with an interfacial DM interaction.\cite{Zhang-SciRep2015,Heo-SciRep2016} We also provide a (mean-field) phase diagram parametrized by the coupling constants of the competing DM terms and show that in-plane skyrmions occupy a substantial area of it, which represents our second main finding. We also study the effect of dipolar interactions on the stability of these spin textures (Sec.~II B). Sec.~III deals with the creation of in-plane skyrmions by means of two different mechanisms, namely, blowing of in-plane domain wall (DW) pairs through a geometric constriction (Sec.~III A) and shedding of skyrmions from a magnetic inhomogenetity (Sec.~III B). In Sec.~IV we explore the dynamical features of in-plane skyrmions driven by spin-transfer and spin-orbit torques, both analytically and micromagnetically, and compare our findings to those arising in the N\'{e}el scenario. This constitutes the third main finding of our work. Sec.~V concerns analytical aspects of the phase diagram depicted in Sec.~II.  In Sec.~VI, we present an exactly solvable model for skyrmions with an interfacial-like DM term in the absence of stray fields and for an arbitrary direction of the background magnetization (in a plane perpendicular to the film), from which in-plane skyrmions emerge as particular solutions. The exact solvability of the model lies in it being defined at critical coupling, a requirement originally introduced in Ref.~\onlinecite{Barton-2018} where the properties of exactly solvable models for skyrmions with a bulk DM term were studied. We note that our exact solution results for the critical coupling model have been recently reported 
in Ref.~\onlinecite{Schroers-2019} within the framework of gauge nonlinear sigma models defined on arbitrary Riemann surfaces. Furthermore, in Sec.~VII we discuss our results and provide concluding remarks. We complement this manuscript with four Appendices containing mathematical identities and further derivations of the aforementioned analytical results.

\section{Minimal model and crystal symmetries}

\begin{table*}[ht!]
\begin{center}
\label{Tab1}
  \begin{tabular}{ c | c | c | c | c | c |}
  \cline{2-6}
  & $A$ (J/m) & $K$ (J/m$^{3}$) & $K_{\textrm{inh}}$ (J/m$^{3}$) & $M_{s}$ (A/m) & $D$ (J/m$^{2}$)\\
     \hline
   \multicolumn{1}{|c|}{\textrm{Phase diagram}}& $3\cdot10^{-11}$ & $5.0\cdot10^{5}$ & $0$ & $5.8\cdot10^{5}$ & \textrm{variable}\\ \hline
    \multicolumn{1}{|c|}{\textrm{Production via blowing}}& $3\cdot10^{-11}$ & $3.0\cdot10^{5}$ & $0$ & $5.8\cdot10^{5}$ & $2.5\cdot10^{-3}$\\ \hline
     \multicolumn{1}{|c|}{\textrm{Production via shedding}}& $4.2\cdot10^{-11}$ & $8.0\cdot10^{4}$ & $8\cdot10^{4}$ & $3.0\cdot10^{5}$ & $1.2\cdot10^{-3}$\\ \hline
      \multicolumn{1}{|c|}{\textrm{Current-driven dynamics}}& $3\cdot10^{-11}$ & $5.0\cdot10^{5}$ & $0$ & $5.8\cdot10^{5}$ & $3.0\cdot10^{-3}$\\ \hline
 \end{tabular}
  \caption{Values of the micromagnetic parameters utilized in the simulations. In all figures we considered single in-plane DMI, see Eq.~\eqref{E_functional_in-plane}, except for the phase diagram of the monoclinic $Cm$ system, in which we used two different Dzyaloshinskii coupling constants, see Eq.~\eqref{eq:DM_Cm}.}
  \label{Table1}
\end{center}
\end{table*}

We consider thin films made of a collinear ferromagnet deposited on a (heavy-metal) substrate. The heterostructure is assumed to be magnetically uniform along the vertical ($z$-axis) direction, which occurs e.g.\ when the film thickness is less than the exchange length of the ferromagnetic material. We start by constructing a minimal model for a two-dimensional ferromagnet hosting stable in-plane skyrmions in a uniform magnetization background along the $x$ direction. The geometry of the heterostructure is defined by the surface $\mathcal{S}$ in the $xy$-plane. 
It is worth remarking that an in-plane skyrmion configuration can be obtained by rotating an out-of-plane skyrmion (magnetization background pointing along the normal to the ferromagnet) by $90\degree$ around an in-plane axis. In particular, the energy functional for the former can be obtained from that corresponding to the latter by applying a rotation in spin space of angle $\alpha=-\pi/2$ along the $y$ axis:\cite{Gobel-PRB2019}
\begin{align}
\label{E_functional_in-plane}
\mathcal{E}_{\textrm{in-pl}}[\bm{m}]=\int_{\mathcal{S}}d&^{2}\vec{r}\,\bigg\{\frac{A}{2}(\vec{\nabla}\bm{m})^{2}+D\mathcal{L}_{\textrm{in-pl}}[\bm{m}]+\nonumber\\
&K[1-(\bm{m}^{x})^{2}]-\bm{m}\cdot(\bm{H}+\bm{H}_{\textrm{d}})\bigg\},
\end{align}
where $\bm{m}$ denotes the order parameter of the ferromagnet, $|\bm{m}|=1$, $\bm{H}$ and $\bm{H}_{\textrm{d}}$ are the normalized (by the saturation magnetization $M_{s}$) external and dipolar magnetic fields, respectively, and 
\begin{equation}
\mathcal{L}_{\textrm{in-pl}}[\bm{m}]=\bm{m}^{z}\partial_{x}\bm{m}^{x}-\bm{m}^{x}\partial_{x}\bm{m}^{z}+\bm{m}^{x}\partial_{y}\bm{m}^{y}-\bm{m}^{y}\partial_{y}\bm{m}^{x}
\end{equation}
is a Lifshitz invariant. Here, $A$, $D$ and $K$ are the exchange stiffness, Dzyaloshinskii and on-site anisotropy constants of the system. The model is minimal in the sense that it contains all symmetry-allowed exchange, Zeeman, dipolar and relativistic energy terms up to second order in the magnetization vector and its spatial derivatives. Hereafter we will use lower indices and vector arrows for real space and upper indices and bold symbols for spin-space
in what follows.

\begin{figure}
\begin{center}
\includegraphics[width=1.0\columnwidth]{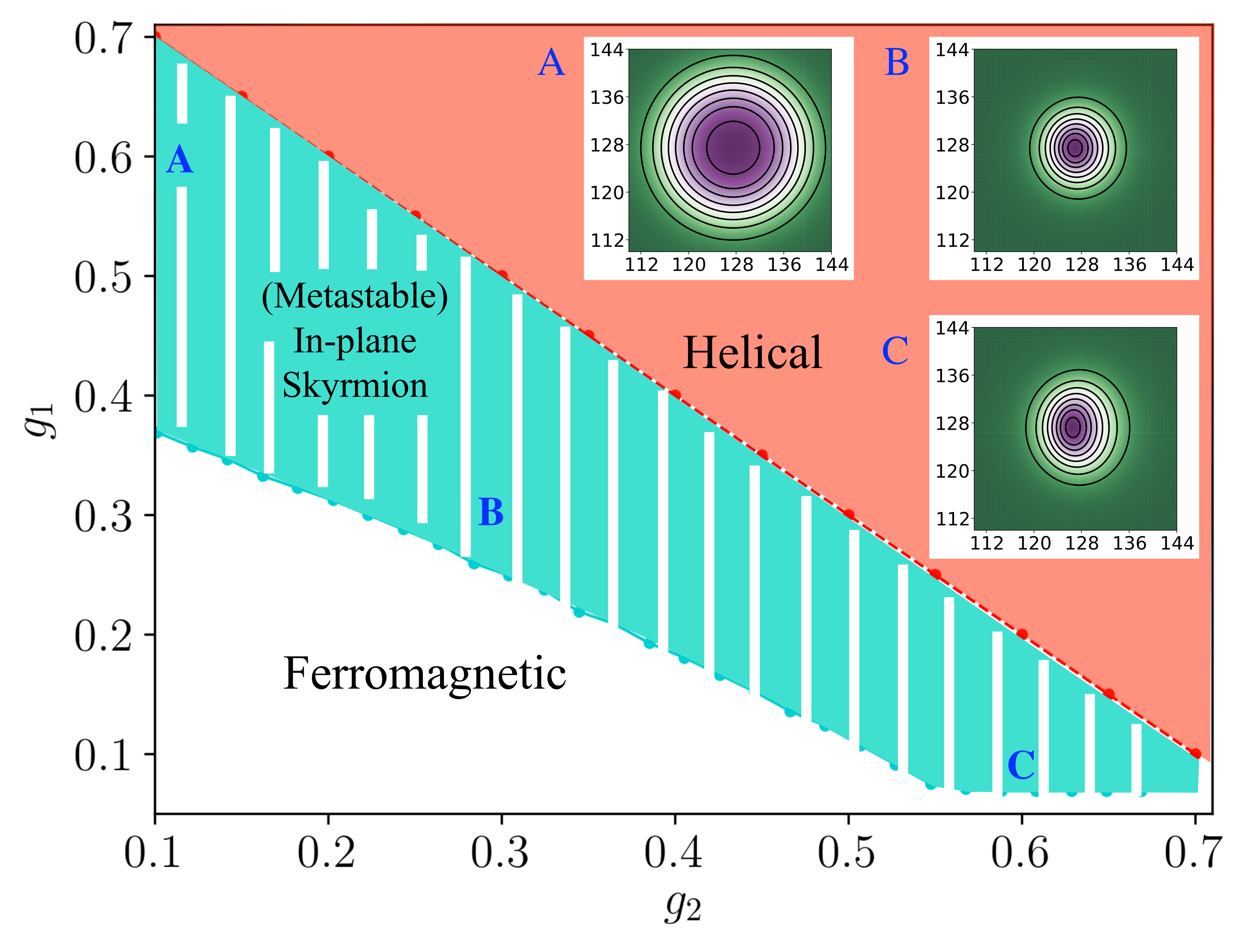}
\caption{Phase diagram of a monoclinic $Cm$ magnet as a function of the in-plane ($g_{1}$) and interfacial ($g_{2}$) reduced Dzyaloshinskii coupling constants. The ferromagnetic phase and helical phase boundary is depicted by the red dashed line. Within the ferromagnetic phase, we find a region of metastable isolated in-plane skyrmions (indicated in light blue). Examples of skyrmion shapes are shown in the insets corresponding to the A,B, and C points in the phase diagram. Increasing the strength of the interfacial DMI deforms the skyrmion.
}
\vspace{-0.5cm} 
\label{Fig1}
\end{center}
\end{figure}

\subsection{Crystal symmetries and materials}
In general, the functional form of the Dzyaloshinkii-Moriya (DM) energy reads $\mathcal{E}_{\textrm{DM}}[\bm{m}]=D_{ik}^{j}\bm{m}^{i}\partial_{j}\bm{m}^{k}$, where $\hat{D}$ is a third-rank antisymmetric polar tensor.\cite{LL} Note that Eq.~\eqref{E_functional_in-plane} yields the following nonzero coefficients of the DM tensor, $D_{zx}^{x}=D_{xy}^{y}=-D_{xz}^{x}=-D_{yx}^{y}=D$ for the minimal in-plane skyrmion model. According to Neumann's principle, the structure of the DM tensor is governed by the crystal symmetry of the ferromagnet:\cite{Powell-Book} $\hat{D}$ must be invariant under the action of all symmetry operations $R$ in the point group, whose components transform as $D_{i'k'}^{j'}=R_{i'i}R_{j'j}R_{k'k}D_{ik}^{j}$. A thorough symmetry analysis reveals that the monoclinic point-group $Cm$, generated by the mirror-symmetry $\bf{m}_{y}$,
is the only space group compatible with the DM tensor structure of Eq.~\eqref{E_functional_in-plane}: invariance of $\hat{D}$ under $\bf{m}_{y}$ yields the non-vanishing elements $D_{xz}^{x}=-D_{zx}^{x}$, $D_{yz}^{y}=-D_{zy}^{y}$ and $D_{yx}^{y}=-D_{xy}^{y}$ for the DM tensor. The corresponding DM energy functional can be cast as 
\begin{align}
\label{eq:DM_Cm}
\mathcal{E}_{\textrm{DM}}[\bm{m}]&=D_{zy}^{y}\mathcal{L}_{\textrm{N\'{e}el}}[\bm{m}]+D_{xy}^{y}\mathcal{L}_{\textrm{in-pl}}[\bm{m}]\\
&\hspace{1.5cm}+D_{3}(\bm{m}^{z}\partial_{x}\bm{m}^{x}-\bm{m}^{x}\partial_{x}\bm{m}^{z}),\nonumber
\end{align}
where $D_{3}=D_{zx}^{x}-D_{xy}^{y}-D_{zy}^{y}$. Here, $\mathcal{L}_{\textrm{N\'{e}el}}[\bm{m}]=\bm{m}^{z}(\nabla\cdot\bm{m})-(\bm{m}\cdot\nabla)\bm{m}^{z}$ is the Lifshitz invariant enabling the stabilization of N\'{e}el skyrmions. Therefore, monoclinic systems described by the point group $Cm$ exhibit competing twisting interactions at the bulk level.

{\it Ab initio} calculations\cite{Database} predict that materials such as Fe(BRh$_{2}$)$_{3}$, Co(BRh$_{2}$)$_{3}$, FeLa$_{3}$S$_{6}$, Al$_{18}$Co$_{5}$Ni$_{3}$, Rb$_{6}$Fe$_{2}$O$_{5}$, La$_{4}$TaCo$_{33}$, Co$_{25}$Cu$_{11}$O$_{48}$, Ta$_{12}$Co$_{3}$Pt$_{3}$Se$_{32}$ and Li$_{4}$Fe$_{3}$Ni$_{3}$(TeO$_{8}$)$_{2}$ exhibit the $Cm$ space group symmetry. In this regard, we propose FeLa$_{3}$S$_{6}$ and Rb$_{6}$Fe$_{2}$O$_{5}$ as the most feasable platforms to host in-plane skyrmions, since the bulk synthesis of these two materials has been previously reported.\cite{Collin-ACB1974,Villars2016}

Figure~\ref{Fig1} depicts a phase diagram of a thin monoclinic $Cm$ system, disregarding dipolar interactions, as a function of the two reduced Dzyaloshinskii coupling constants $g_{1}=\pi D_{xy}^{y}/4\sqrt{AK}$ and $g_{2}=\pi D_{zy}^{y}/4\sqrt{AK}$ (see Section~\ref{Sec_Analytics} for further details), in which we observe a (uniform) ferromagnetic phase, a helical phase and, within the ferromagnetic phase, a region where single in-plane skyrmions are metastable. Note that only the first two phases correspond to true ground states of the magnet, whereas skyrmions emerge as low-energy excitations in the magnetic vacuum (uniform background). The linear boundary between uniform and helical phases has been calculated analytically (see Section~\ref{Sec_Analytics} for a description of the techniques employed).
 The term proportional to $D_{3}$ has been disregarded in our analysis since it is only responsible for the ellipticity (shape deformation) of the resultant skyrmions. Whether in-plane skyrmions are stabilized in this class of magnetic materials for an in-plane (easy $x$-axis) anisotropy in the absence of external magnetic fields and dipolar interactions depends on the interplay between N\'{e}el and in-plane DM stabilizers. Micromagnetic simulations to support our analytic predictions were performed in Mumax3,\cite{Mumax} in a square geometry of lateral size 256 nm and thickness 1 nm, with cell size $1\times 1\times 1$ nm$^{3}$ disregarding dipolar interactions. The values for the micromagnetic parameters are given in Table~\ref{Table1}. The phase diagram was calculated by sweeping $g_{1}$ and $g_{2}$ within the range $0-1$. The (blue) domain of skyrmion metastability was obtained by studying, for the set of parameters corresponding to the uniform ferromagnetic phase, whether an initial single in-plane skyrmion configuration relaxed towards the uniform state or not. 

\subsection{Effect of dipolar interactions on skyrmions}

\begin{figure}
\begin{center}
\includegraphics[width=1.0\columnwidth]{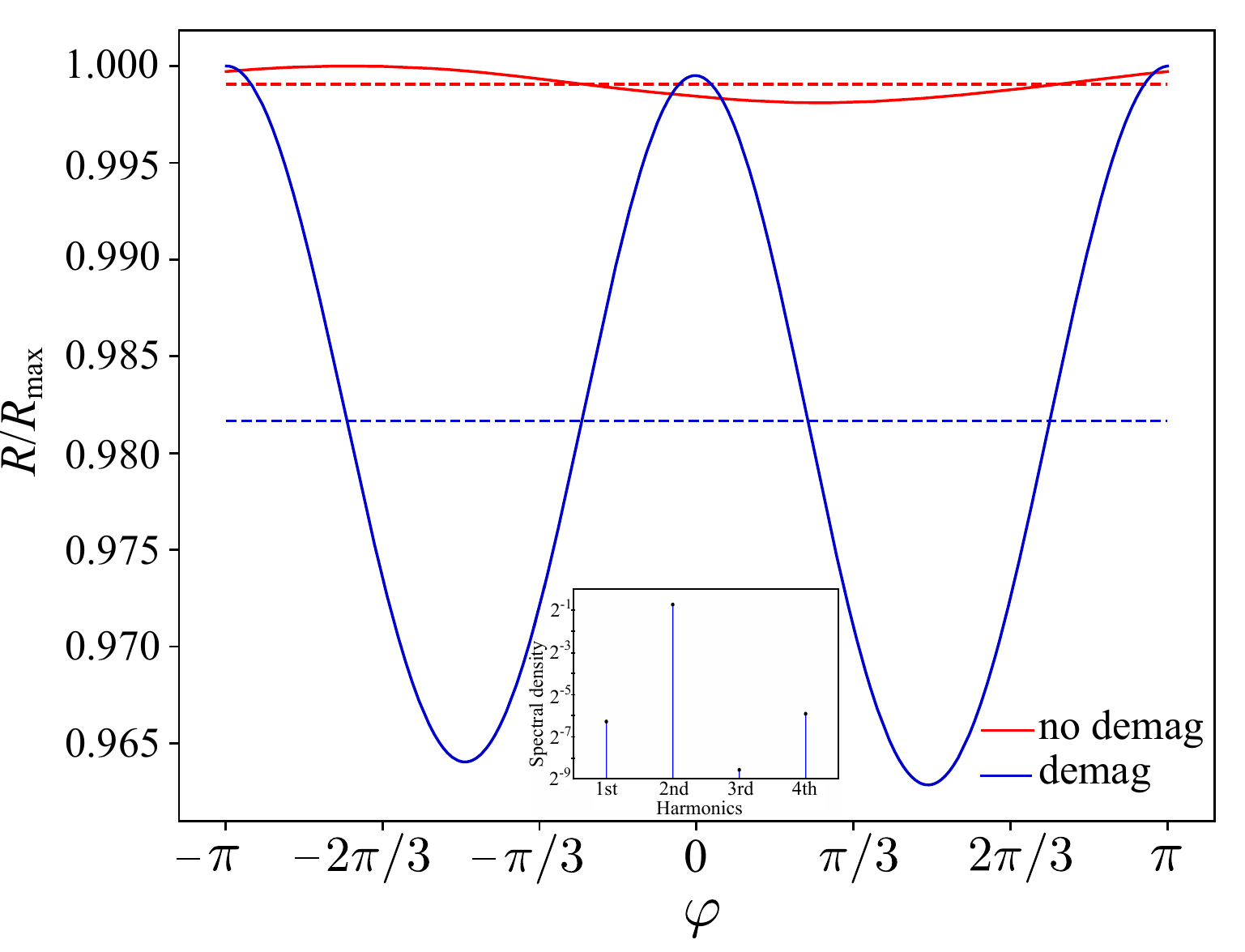}
\caption{Radius of in-plane skyrmions as a function of the azimuthal angle $\varphi$ in the presence (blue) and absence (red) of dipolar interactions (demag). Dashed lines represent the angular average (mean) of the radius. We have normalized the the radii by the maximal radius $R_{\textrm{max}}$ observed for the skyrmion. Note that the numerical estimate of the radius is affected by finite size effects on the order of the cell size ($\sim1$nm), therefore the red curve corresponds to the average. (Inset) Spectral density corresponding to the aforementioned angular dependence for the stray-field case.}
\vspace{-0.5cm} 
\label{Fig2}
\end{center}
\end{figure}

The effect of magnetostatic interactions cannot be studied by just performing the rotation in spin-space starting from the N\'eel skrymion. In this subsection, we show that their influence is stronger for in-plane skyrmions and lead to the deformation of the latter. Figure~\ref{Fig2} shows the angular dependence of the radius for in-plane skyrmions in the presence and absence of stray fields, with the radius $R(\varphi)$ being defined by the condition $\bm{m}^{x}[R(\varphi),\varphi]=0$ (i.e., the position at which $\bm{m}^{x}$ vanishes) and $\varphi$ denoting the azimuthal angle.

In contrast to the N\'{e}el scenario, we observe oscillations in $R$ when $\varphi$ is swept within the range $[0,2\pi]$, in the presence of  dipolar interactions. In what follows, we refer to the angular average (mean) of $R$ as the {\it skyrmion radius}, although this definition is only meaningful in the absence of stray fields since radial fluctuations are of the order of $0.2\%$ in this case. Furthermore, we have also calculated the spectral density of $R(\varphi)$ in the presence of demagnetizing fields, which we depict in the inset of Fig.~\ref{Fig2}; we obtain the harmonic terms up to fourth order in the natural frequency $\omega_{\varphi}=1$ and conclude that the dominant contribution stems from the second-order one. 

\begin{figure}
\begin{center}
\includegraphics[width=1.0\columnwidth]{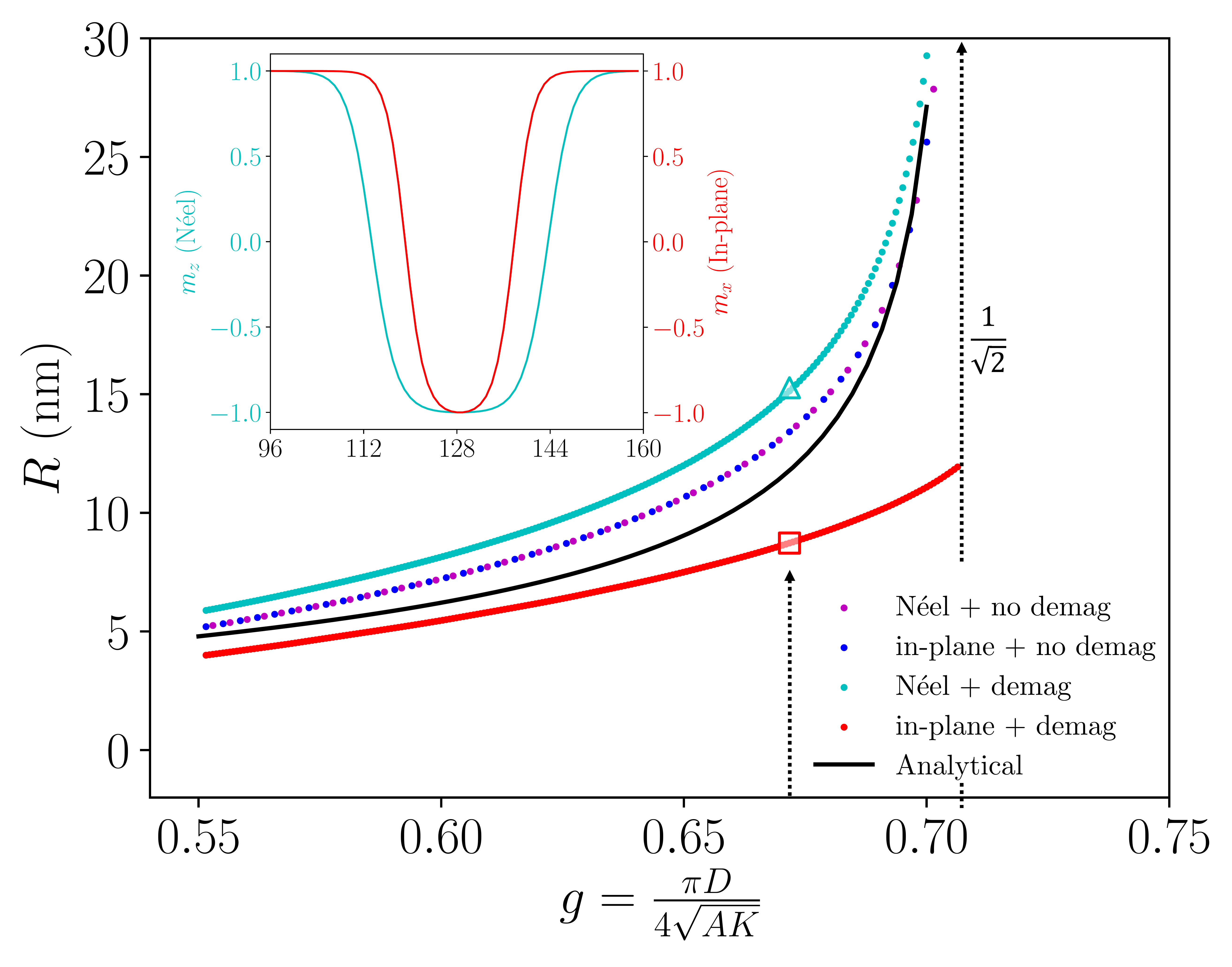}
\caption{Radius of N\'{e}el and in-plane skyrmions in the presence (blue and magenta curves) and absence (cyan and red curves) of dipolar interactions (demag). The black solid curve denotes the analytical dependence on the reduced DM parameter $g=\pi D/4\sqrt{AK}$ proposed in Ref.~\onlinecite{Kravchuk2018}. (Inset) Radial magnetization profile $\bm{m}^{z}$ (red) and $\bm{m}^{x}$ (blue) in the presence of stray fields for N\'{e}el and in-plane skyrmions, respectively, where the corresponding radii are marked by \textcolor{red}{$\square$} and \textcolor{cyan}{$\Delta$} in the $R-g$ curves. The dashed black arrow indicates the associated $g$-value.}
\vspace{-0.5cm} 
\label{Fig3}
\end{center}
\end{figure}

The effect of dipolar fields on the stability of skyrmion textures emerging in the proposed quasi-two-dimensional geometry has been studied by performing micromagnetic simulations in Mumax3 and Micromagnum. \cite{Micromagnum} We have considered both in-plane and N\'{e}el skyrmions, the latter being soliton solutions of the energy functional given by Eq.~\eqref{E_functional_in-plane} with account of the substitutions $\mathcal{L}_{\textrm{in-pl}}[\bm{m}]\rightarrow\mathcal{L}_{\textrm{N\'{e}el}}[\bm{m}]$ and $1-(\bm{m}^{x})^{2}\rightarrow1-(\bm{m}^{z})^{2}$ in the DM and anisotropy terms, respectively. Simulations were carried out in a square geometry of lateral size 512 nm and thickness 1 nm, with cell size $1\times 1\times 1$ nm$^{3}$; we took the same values for the micromagnetic parameters as those used in the calculation of the phase diagram, see Fig. \ref{Fig1}. The skyrmion radius $R$ is, again, extracted from the isocontours $\bm{m}^{x}=0$ (in-plane) and $\bm{m}^{z}=0$ (N\'{e}el) obtained after relaxation of an initial magnetic configuration consisting of a disk of radius 64 nm with uniform magnetization along the opposite direction to that of the uniform magnetic background ($z$ axis for the N\'eel case), see inset of Fig.~\ref{Fig3}.\cite{FN1} Note that here we do consider skyrmions with a small domain wall (DW) width compared to their radius $R$. The dependence of the skyrmion radius on the reduced DM parameter $g=\pi D/4\sqrt{AK}$ is depicted in Fig.~\ref{Fig3} in the absence of external magnetic fields for both N\'{e}el and in-plane skyrmions. We observe that in the absence of dipolar interactions the $R$-$g$ curves overlap, which results from the fact that, in this case, the corresponding energy functionals can be mapped into each other via a rotation in spin space. When stray fields are accounted for, however, the skyrmion radius increases for the N\'{e}el configuration and decreases for the in-plane one. This behavioural difference originates in the fact that dipolar fields are weaker for N\'{e}el skyrmions, since the contribution of the magnetic charge, $\nabla\cdot\bm{m}$, can be effectively disregarded in this case.\cite{FN2} Within the range  $0.35\leq g\leq0.7$, the skyrmion radius is observed to be less than 15 nm for in-plane skyrmions, in stark contrast to the N\'{e}el case (range $5-40$ nm). Finally, the black curve depicts the analytical solution  $R=\sqrt{2}|g|/\sqrt{1-2g^{2}}$ derived in Ref.~\onlinecite{Kravchuk2018} for comparison.

\begin{figure}
\begin{center}
\includegraphics[width=1.0\columnwidth]{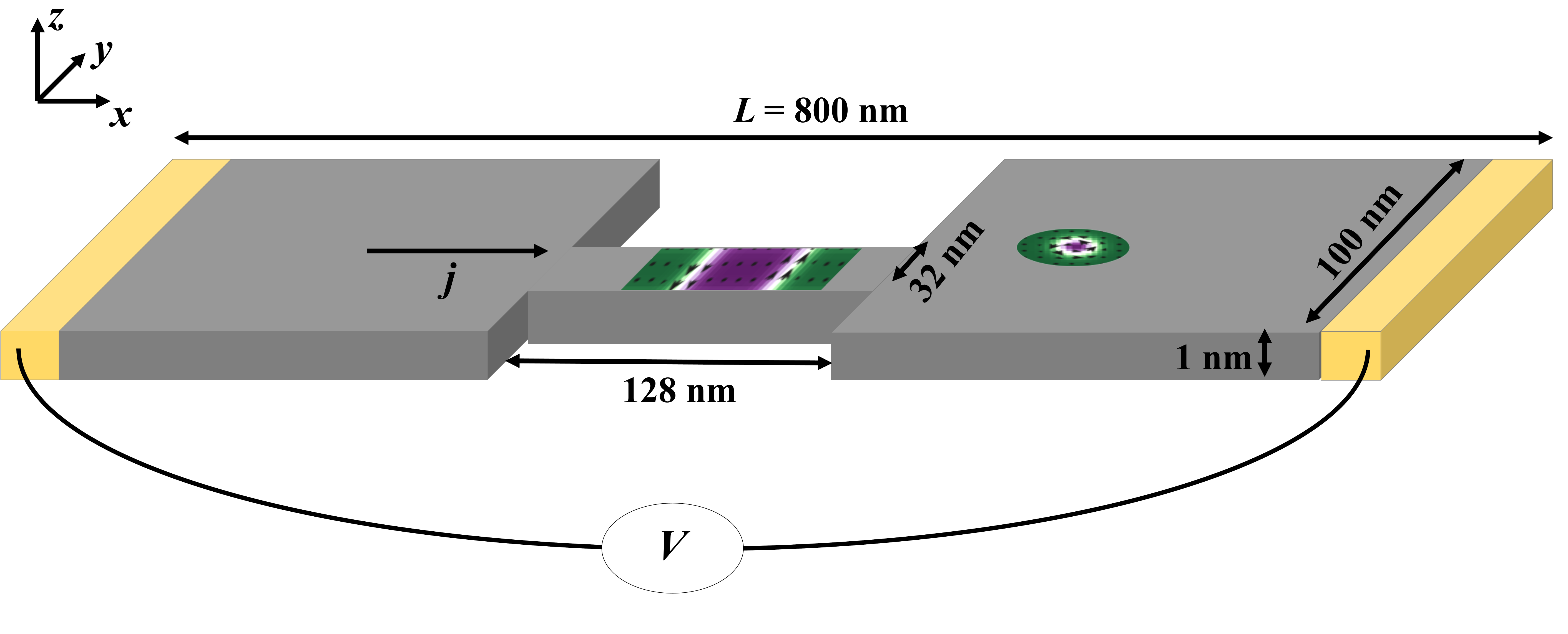}
\includegraphics[width=1.0\columnwidth]{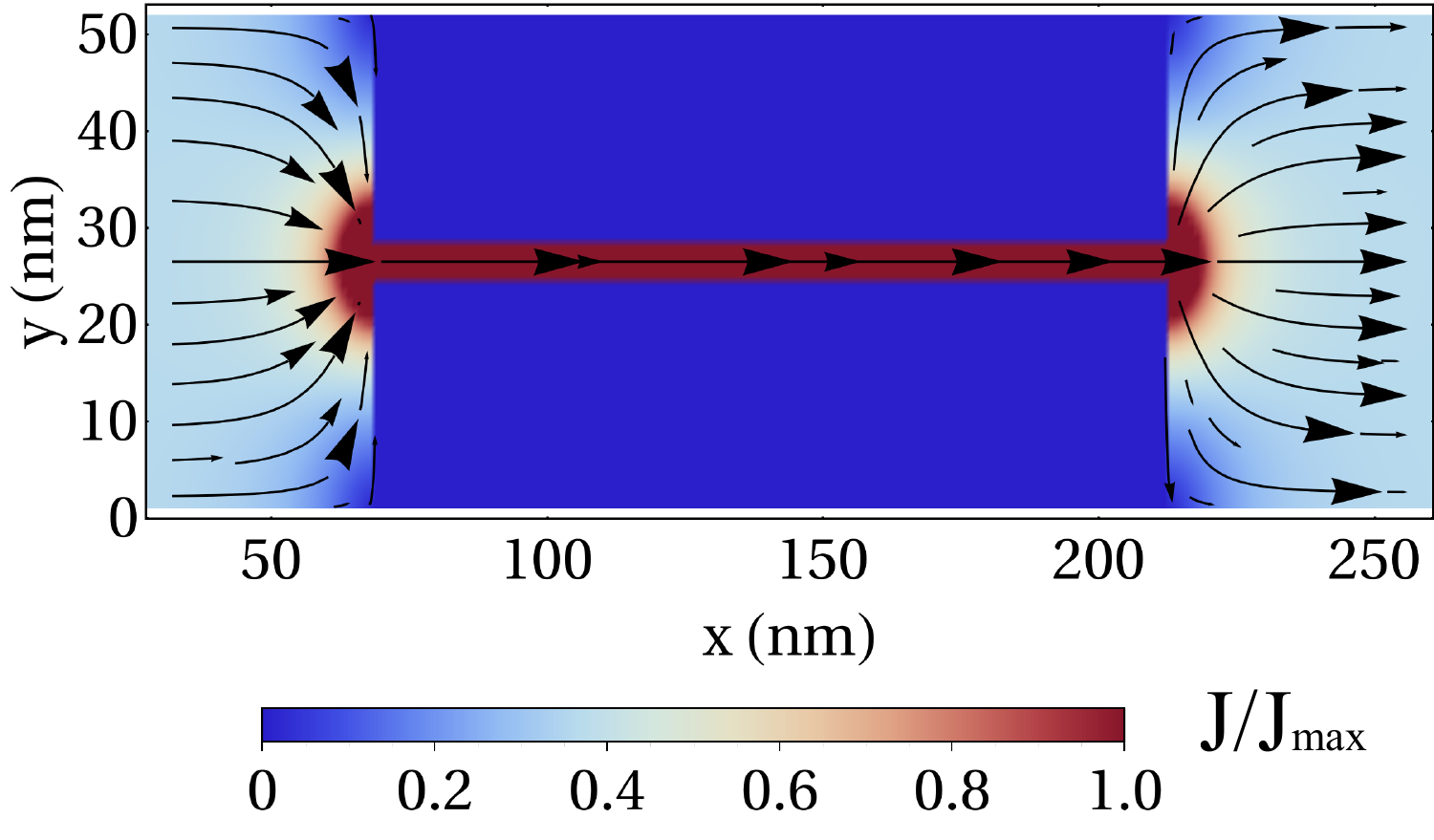}
\caption{(a) Sketch of the geometry utilized in the numerical simulations of the production of in-plane skyrmions (green-white disk) via the blowing method. (b) Spatial distribution of the electric-current density calculated self-consistently with constant-voltage boundary conditions (see main text). $J_{\textrm{max}}=2\phi|_{\text{left}}/L$ denotes the maximum current injected into the sample of length $L$.}
\vspace{-0.5cm} 
\label{Fig4}
\end{center}
\end{figure}

\section{Skyrmion production}
In this section we show that in-plane skyrmions can be produced in an analog way to N\'eel skyrmions despite the different influences of dipolar effects. In the first part we present on how to create them by blowing magnetic bubbles through a geometrical constriction\cite{Jiang-Sci2015} and in the second part we show how to create them via an interplay of a magnetic inhomogeneity and a current.\cite{Sitte-PRB2016,Everschor-Sitte-NJP2017,Stier-PRL2017,Buttner-NNano2017}

\begin{figure*}
\begin{center}
\includegraphics[width=0.8\linewidth]{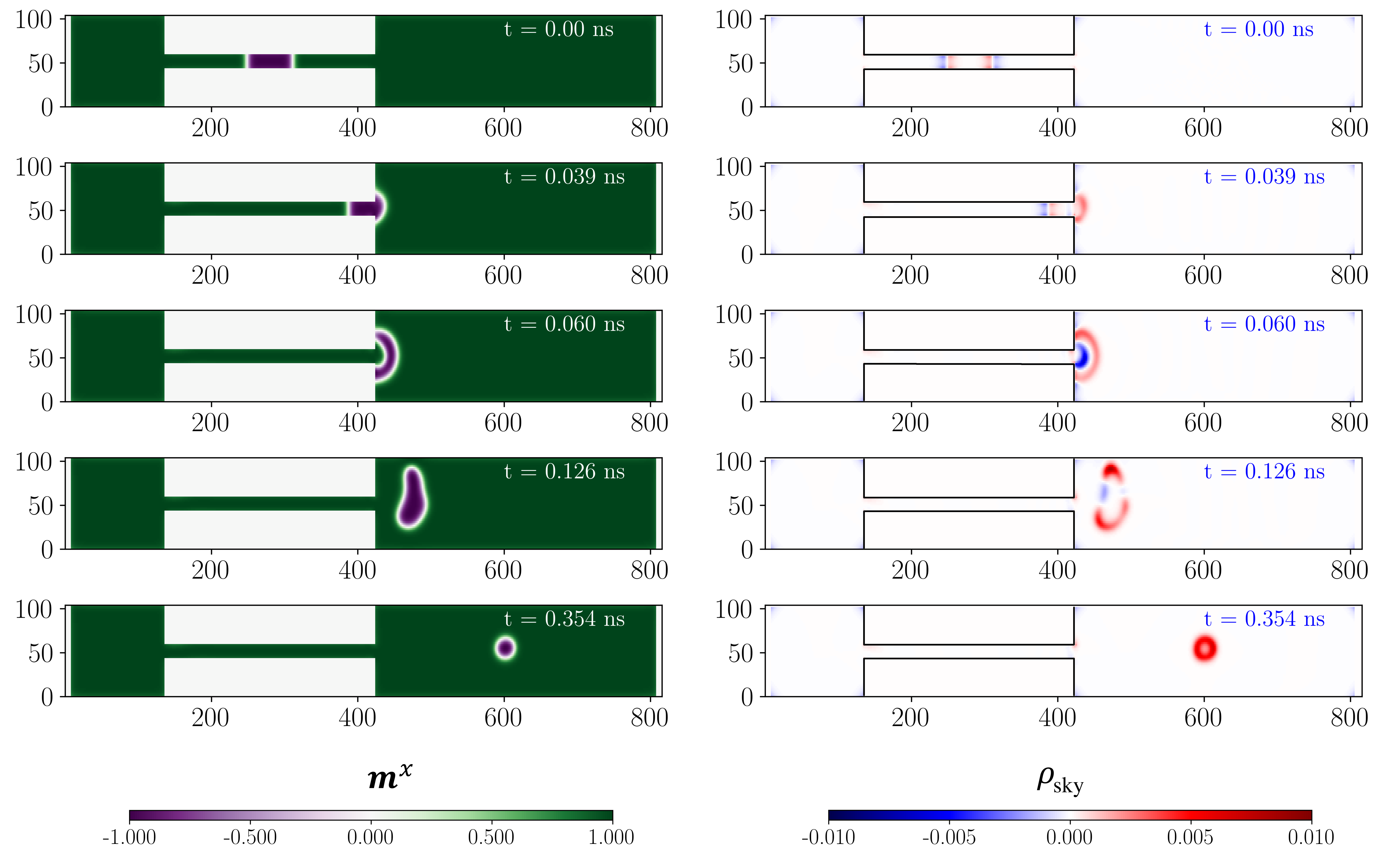}
\caption{Generation of in-plane skyrmions via the current-driven motion of a DW pair through a narrow geometrical constriction. (Left panel) Snapshots of the $x$-component of the order parameter, $\bm{m}^{x}$, taken at five sequential times. (Right panel) Time evolution of the skyrmion charge density $\rho_{\textrm{sky}}$ corresponding to the magnetization texture depicted on the left panel. The color code indicates the $\bm{m}^{x}$ component of the magnetization.}
\vspace{-0.5cm} 
\label{Fig5}
\end{center}
\end{figure*}

\subsection{Blowing in-plane skyrmions}
In-plane skyrmions can be produced by blowing magnetic bubbles through a geometrical constriction, akin to the N\'{e}el case.\cite{Jiang-Sci2015} Numerical simulations have been carried out in Micromagnum with extensions to directly solve for the current flow.\cite{Micromagnum} The geometry considered was a thin magnetic film of dimensions 800$\times$100$\times$1 nm$^{3}$ (length$\times$width$\times$thickness), which presents a narrow constriction of length 128 nm at its center, see Fig.~\ref{Fig4}(a). The values of the micromagnetic parameters utilized can be found in Table~\ref{Table1}. An electromotive force is applied longitudinally between the metal contacts deposited on the lateral sides of the magnet; the charge-current density is self-consistently calculated\cite{Micromagnum} by solving, at each time step, the Poisson equation $\nabla^{2}\phi=0$ with boundary conditions $\phi|_{\text{left}}=-\phi|_{\textrm{right}}=2.55$ V in the proposed geometry. The current density is obtained from $\vec{j}=-\hat{\sigma}\vec{\nabla}\phi$, where $\hat{\sigma}$ is the conductivity tensor,\cite{FN3} see Fig.~\ref{Fig4}(b).

Time evolution of the order parameter obeys the Landau-Lifshitz-Gilbert (LLG) equation
\begin{equation}
\label{LLG}
\partial_{t}\bm{m}=-\gamma\,\bm{m}\times\bm{H}_{\textrm{eff}}+\alpha\,\bm{m}\times\partial_{t}\bm{m}+\bm{\tau},
\end{equation}
where $\bm{H}_{\textrm{eff}}=-\frac{1}{M_s}\delta_{\bm{m}}\mathcal{E}_{\textrm{in-pl}}[\bm{m}]$ is the thermodynamic force conjugate to $\bm{m}$, $\alpha$ is the Gilbert damping constant and $\bm{\tau}$ denotes the
current-induced magnetic torque. In our simulations, the initial magnetic configuration consists of a DW pair located at the center of the constriction, whose dynamics are driven by the spin-transfer torque  $\bm{\tau}_{\textrm{STT}}=-\zeta(\beta+\bm{m}\times)\bm{m}\times(\vec{j}\cdot\vec{\nabla})\bm{m}$. Here, $\zeta=\gamma P \hbar/2e M_{s}$ is the charge-to-spin conversion factor in the adiabatic regime, $\gamma$ and $P$ represent the gyromagnetic ratio and the electron spin polarization, respectively, and $\beta$ parametrizes the effect of spin-dephasing processes on the transfer of angular momentum between the electron flow and the magnet. As a result of the current-driven motion, the DW pair begins to deform and is pushed out towards the right end of the channel, eventually detaching from it. Outside the constriction, the magnetic texture evolves into an in-plane skyrmion and generically performs a curved trajectory towards one of the transversal sides of the film due to the Magnus force. The production process just described is depicted in Fig.~\ref{Fig5}, where we show snapshots of the DW pair-to-skyrmion conversion at different times (left panel) and the corresponding topological charge density (right panel), defined by the integrand of the Pontryagin index
\begin{equation}
\label{Sky_charge}
Q=\frac{1}{4\pi}\int_{\mathcal{S}} d^{2}\vec{r}\hspace{0.15cm}\bm{m}\cdot(\partial_{x}\bm{m}\times\partial_{y}\bm{m})=\int_{\mathcal{S}} d^{2}\vec{r}\hspace{0.15cm}\rho_{\textrm{sky}}.
\end{equation}
Note that we restricted ourselves to the case $\alpha=\beta$ to avoid the deflection of the skyrmion trajectories along the transverse ($y$ axis) direction.

\begin{figure}
\begin{center}
\includegraphics[width=0.9\columnwidth]{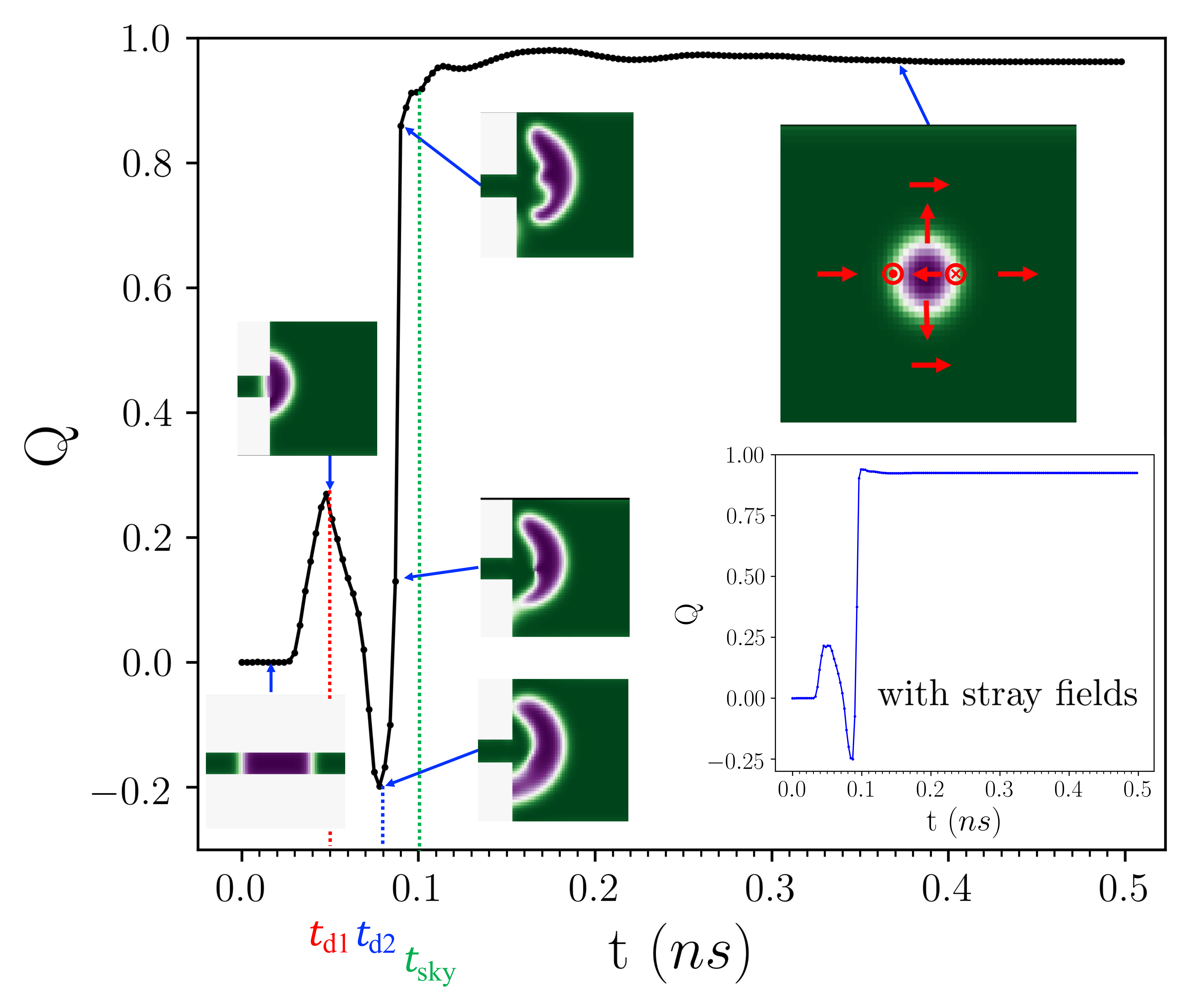}
\caption{Time evolution of the topological charge for the production of in-plane skyrmions via blowing of magnetic bubbles through a geometrical constriction for the value $g=0.523$. In this simulation, corresponding to the geometry of Fig.~\ref{Fig5}, the values of the times $t_{d1}$, $t_{d2}$ and $t_{\textrm{sky}}$ (see main text for the definitions) are 0.04 ns, 0.08 ns and 0.17 ns respectively. (Inset) Time evolution of the topological charge in the presence of dipolar interactions. The corresponding values of the characteristic times are $t_{d1}=0.05$ ns, $t_{d2}=0.85$ ns and $t_{\textrm{sky}}=0.1$ ns.}
\vspace{-0.5cm} 
\label{Fig6}
\end{center}
\end{figure}

Figure~\ref{Fig6} shows the temporal dependence of the skyrmion charge during the production process of in-plane skyrmions. $Q$ has been calculated at every time step of the simulation by integrating the skyrmion charge density depicted in the right panel of Fig.~\ref{Fig5} over the whole sample. In general, we find that the skyrmion charge first increases monotonically with time and peaks at $t_{d1}$, then decreases to negative values until $t_{d2}$ and, finally, it increases sharply until $t_{\textrm{sky}}$ to the value $Q\simeq1$. From the right panel of Fig.~\ref{Fig5} we observe that $t_{d1}$ corresponds to the time at which the right DW detaches from the right end of the constriction and starts expanding radially, whereas the left DW is still inside the channel. Afterwards, the left DW detaches from the constriction at the time $t_{d2}$ (see the corresponding snapshot in Fig.~\ref{Fig6}) and also expands radially outside the channel, therefore yielding the decrease of the skyrmion charge. Once the DW pair is completely detached from the constriction, it deforms smoothly into an in-plane skyrmion with topological charge $Q\simeq1$ at $t_{\textrm{sky}}$. The skyrmion charge exhibits a qualitatively similar behavior in the presence of dipolar interactions, the main difference being the increase of $t_{d1,d2}$ and the reduction of $t_{\textrm{sky}}$, i.e., it takes a longer time for both DWs to detach from the constriction and a shorter time for the detached magnetic texture to become an in-plane skyrmion due to the action of the stray fields.

\begin{figure}
\begin{center}
\includegraphics[width=0.9\columnwidth]{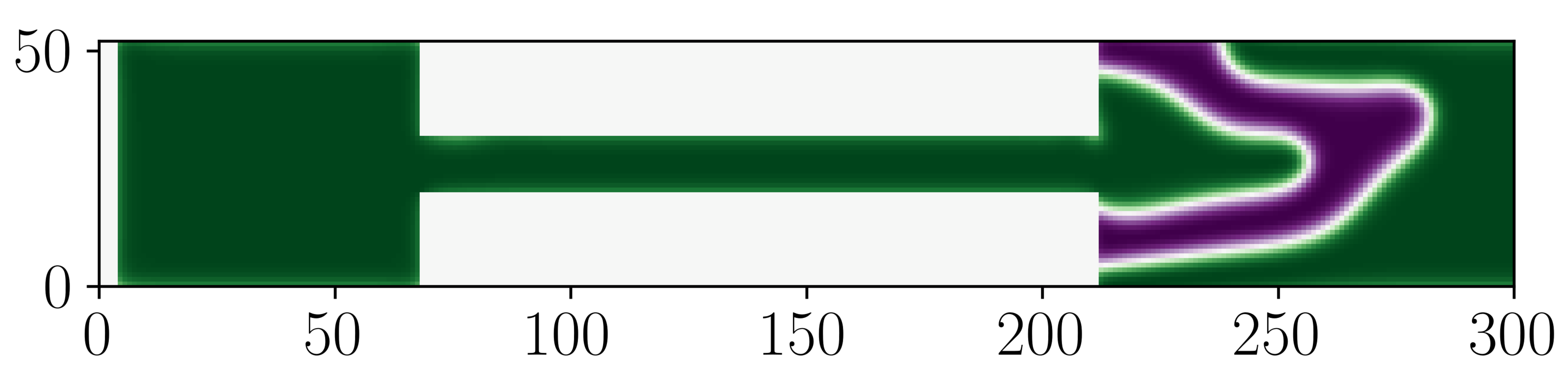}
\caption{Elongated magnetic configuration generated via blowing of a magnetic bubble through a geometrical constriction for the value  $g=0.811$ of the reduced parameter.}
\vspace{-0.5cm} 
\label{Fig7}
\end{center}
\end{figure}

This production scenario for in-plane skyrmions is valid for constriction widths much smaller than that of the sample, $\mathfrak{t}\ll w$. We verified that, in this regime, the production process just described is qualitatively similar to that induced by a stepwise current distribution, uniform inside and outside the constriction, with the inside-to-outside current ratio being proportional to $w/\mathfrak{t}$. On the contrary, if both the constriction and sample widths are comparable, then in-plane skyrmions cannot be produced via magnetic blowing since the DWs just spread over the sample once the injected current pushes them outside the constriction and, therefore, the topology of the magnetic texture remains trivial. Furthermore, we have studied the production of in-plane skyrmions numerically over a wide range of reduced parameters $g=\pi D/4\sqrt{AK}$. We observed that these magnetic textures are produced within the range  $0<g<0.7$, whereas for $g>0.7$ the DW pair becomes attached to the wall outside the channel and then grows into an elongated magnetic configuration, eventually touching the edges of the sample, see Fig.~\ref{Fig7}.

\subsection{Skyrmion creation via an inhomogeneity}

\begin{figure*}
\begin{center}
\includegraphics[width=0.7\linewidth]{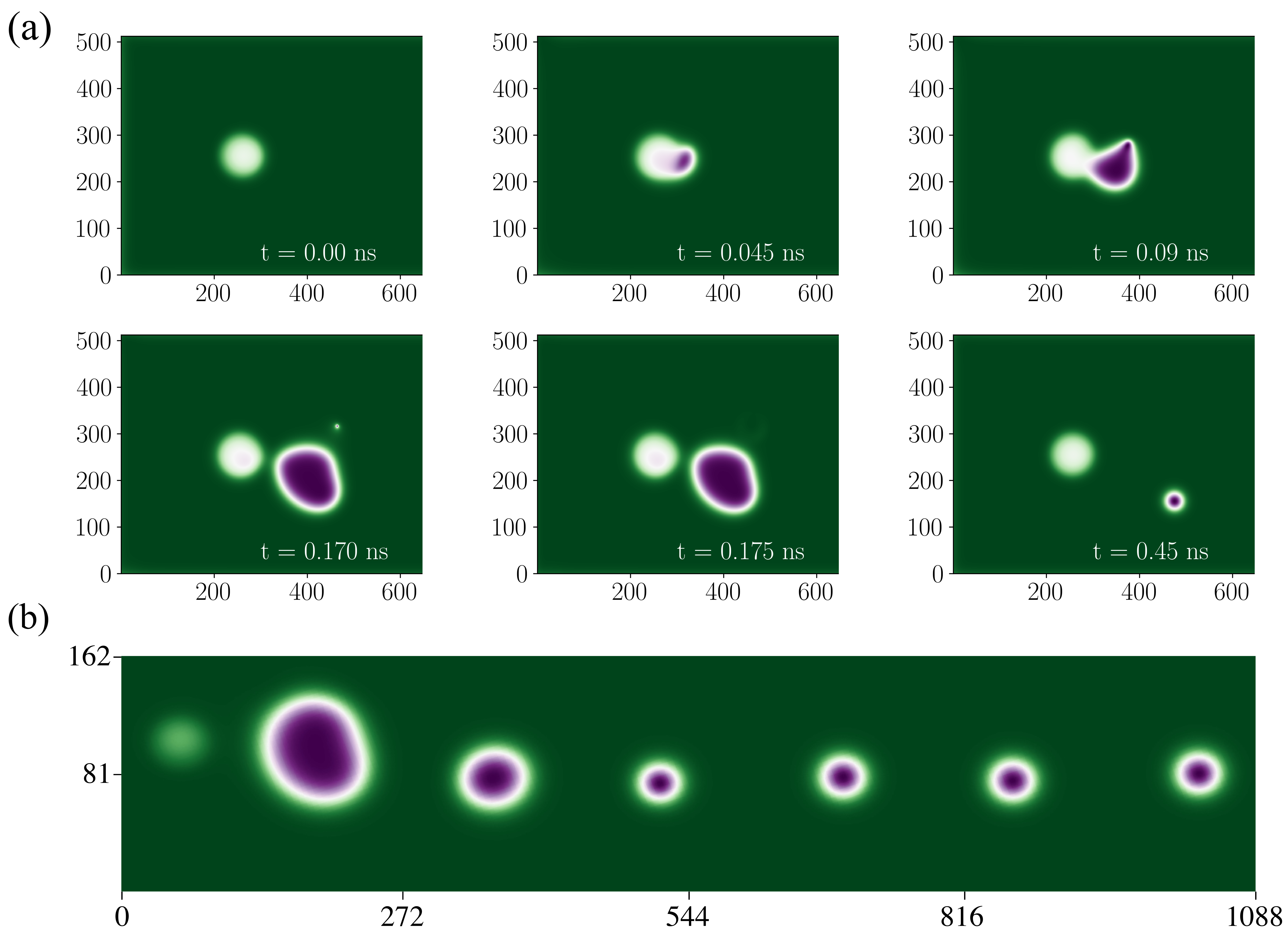}
\caption{Production of in-plane skyrmions via an inhomogeneity. (a) Six snapshots 
illustrating consecutive stages of the shedding process, namely, the initial magnetic configuration, the onset of the magnetic instability, its expansion over the sample, the formation of the skyrmion/antiskyrmion pair, the annihilation of the antiskyrmion and the stabilization of the skyrmion. Full simulation size is 1024$\times$1000 nm$^{2}$. (b) In-plane skyrmion racetrack generated by the shedding method. The color code indicates the $\bm{m}^{x}$ component of the magnetization. Full simulation size is dimensions 1526$\times$1024 nm$^{2}$.}
\vspace{-0.5cm} 
\label{Fig8}
\end{center}
\end{figure*}

Another method for the production of N\'{e}el skyrmions has been recently proposed,\cite{Sitte-PRB2016,Everschor-Sitte-NJP2017,Stier-PRL2017,Buttner-NNano2017} which is based on the (static) loss of stability of a magnetic inhomogeneity triggered by current-induced spin-transfer torques. Here, we exploit the same principle for in-plane skyrmions, showing qualitatively similar results to those of the N\'{e}el case. The initial magnetic configuration is taken to be the uniform state along the easy $x$-axis, except for an inhomogeneous circular domain of radius $50$ nm with magnetization pointing along the $z$ axis, and we applied dc charge currents. Figure~\ref{Fig8} illustrates the results of our simulations in Mumax3 for a dc current $j=7.5\cdot10^{12}$ A/m$^{2}$, where the values of the micromagnetic parameters used can be found in Table~\ref{Table1}: subfig.~(a) depicts details of the shedding process in a geometry of dimensions 1024$\times$1000$\times$0.4 nm$^{3}$. The six (time-)consecutive snapshots correspond to the initial magnetic configuration, the onset of the magnetic instability, its expansion over space, the formation of the skyrmion/antiskyrmion pair, the annihilation of the antiskyrmion and the stabilization of the skyrmion, respectively. 

Subfig.~\ref{Fig8}(b) shows the realization of the racetrack concept for in-plane skyrmions, which are periodically generated via shedding. Here, we are showing a cutout of a larger geometry (of dimensions 1526$\times$1024$\times$0.4 nm$^{3}$) to obtain a larger number of skyrmions within the racetrack.
The value of the critical current for the shedding process is less or approximately $j_{c}\lesssim 3.3\cdot 10^{12}$ A/m$^{2}$ in our simulations. Furthermore, in contrast to the N\'{e}el scenario studied in Ref.~\onlinecite{Everschor-Sitte-NJP2017}, the out-of-plane anisotropy constant $K_{\textrm{inh}}$ associated with the inhomogeneity domain has to be as large as $K$ to guarantee the shedding of in-plane skyrmions, since dipolar interactions favor large in-plane projections of the magnetization; this, in turn, yields larger values of the critical current for shedding.

\section{Current-driven dynamics}

As shown in the previous section, in-plane skyrmions and N\'{e}el skyrmions exhibit similar dynamics driven by spin transfer torques. Therefore, we will focus hereafter on the skyrmion dynamics driven by spin-orbit torques (SOT), which are described by the LLG equation \eqref{LLG} accounting for the torque $\bm{\tau}_{\textrm{SOT}}=(\tau_{\textrm{FL}}+\tau_{\textrm{DL}}\bm{m}\times)(\hat{e}_{z}\times\vec{j}\hspace{0.02cm})\times\bm{m}$. Here, $\tau_{\textrm{FL}}$ and $\tau_{\textrm{DL}}$ parametrize the field-like and damping-like components of the SOT, respectively. Under the assumption of rigidity for the soliton texture, which is valid in the low-frequency (compared to the exchange energy) and low-current regime, we can describe skyrmions by their center of mass $\vec{X}=(X,Y)$. Within the collective-coordinate approach, the dynamics of these solitons obey the following Thiele equation:\cite{Thiele,FN4}
\begin{equation}
\label{Thiele}
G\hat{e}_{z}\times\partial_{t}\vec{X}=\alpha \mathcal{D}\partial_{t}\vec{X}+\tau_{\textrm{FL}}\vec{T}_{\textrm{FL}}+\tau_{\textrm{DL}}\vec{T}_{\textrm{DL}},
\end{equation}
where $G=4\pi Q$ is the gyrotropic constant, $\mathcal{D}$ is the dissipative constant and $\vec{T}_{\textrm{FL}}$ and $\vec{T}_{\textrm{DL}}$ are the forces exerted on the magnetic texture by the field-like and damping-like components of the SOT, respectively. The expressions for these terms read
\begin{align}
\label{Express}
\mathcal{D}&=\frac{1}{2}\int_{\mathcal{S}}d^{2}\vec{r}\hspace{0.15cm}\partial_{k}\bm{m}\cdot\partial_{k}\bm{m},\\
T_{\textrm{FL},i}&=\int_{\mathcal{S}}d^{2}\vec{r}\hspace{0.15cm}(\hat{e}_{z}\times\vec{j}\hspace{0.02cm})\cdot\partial_{i}\bm{m},\nonumber\\
T_{\textrm{DL},i}&=\int_{\mathcal{S}}d^{2}\vec{r}\hspace{0.15cm}\bm{m}\cdot\big[(\hat{e}_{z}\times\vec{j}\hspace{0.02cm})\times \partial_{i}\bm{m}\big],\nonumber
\end{align}
and, by solving the Thiele equation for the skyrmion velocity, we obtain the following expression for the skyrmion Hall angle\cite{Litzius-NatPhys2017}

\begin{figure*}[tb]
\begin{center}
\includegraphics[width=0.7\textwidth]{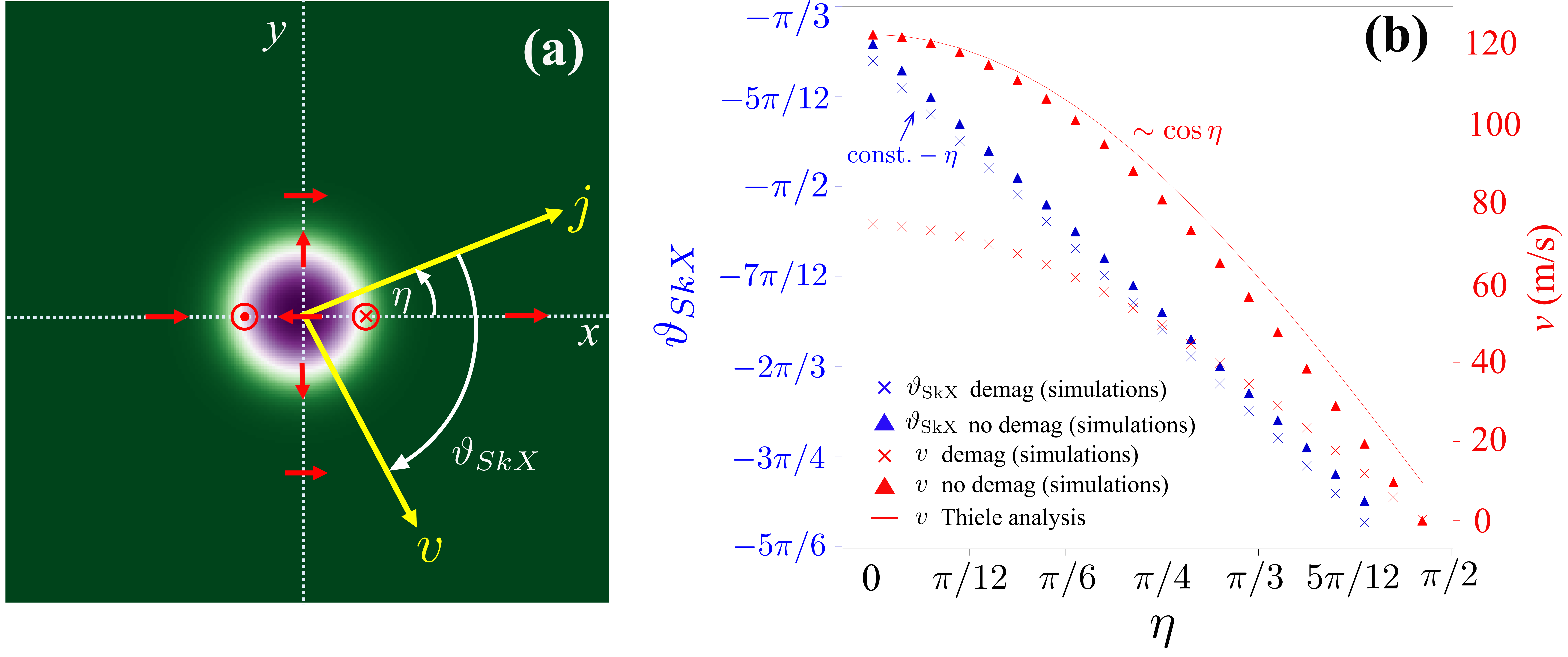}
\caption{Skyrmion Hall effect for the in-plane configuration. (a) Geometry considered for the definition of the skyrmion Hall angle $\vartheta_{SkX}$. Yellow arrows represent the electric current ($\vec{j}$\hspace{0.02cm}) and the skyrmion velocity ($v$), with $\eta$ being the angle between the former and the $x$ axis. Red arrows indicate the direction of the magnetization at the respective points. The color map for the magnetization corresponds to that of Fig.~\ref{Fig4}. (b) Dependence of the skyrmion Hall angle (blue) and terminal speeds (red) on the direction of the injected charge current. The skyrmion velocity has been calculated in the presence (crosses) or absence (triangles) of stray fields, and within the Thiele collective-coordinate approach disregarding stray fields as well (red line).} 
\vspace{-0.5cm} 
\label{Fig9}
\end{center}
\end{figure*}

\begin{equation}
\label{SkyHangle}
\tan(\vartheta_{\textrm{SkX}}+\eta)=\frac{\partial_{t}Y}{\partial_{t}X}=-\frac{GT_{x}-\alpha\mathcal{D}T_{y}}{\alpha\mathcal{D}T_{x}+GT_{y}},
\end{equation}
where $T_{x}\equiv\tau_{\textrm{FL}}T_{\textrm{FL},x}+\tau_{\textrm{DL}}T_{\textrm{DL},x}$, $T_{y}\equiv\tau_{\textrm{FL}}T_{\textrm{FL},y}+\tau_{\textrm{DL}}T_{\textrm{DL},y}$ and $\eta$ is the angle between the injected current and the $x$ axis, as depicted in Fig.~\ref{Fig9}(a). We consider the rigid hard cutoff ansatz for in-plane skyrmions in what follows, which is given by $\bm{m}(\vec{r})=(-\cos\theta(r),\sin\theta(r)\cos\phi,\sin\theta(r)\sin\phi)^{\top}$, with $\vec{r}=r(\cos\phi,\sin\phi)$, $\theta(r)=\pi(1-r/R)\Theta(R-r)$ and $^{\top}$ denoting the transpose operator. Here $\Theta(x)$ is the Heaviside theta function and $R$ denotes the skyrmion radius. By plugging it into Eqs.~\eqref{Express}, we obtain $T_{x}=\pi^{2}jR\tau_{\textrm{DL}}\cos(\eta)/2$ and $T_{y}=0$. We therefore conclude that in-plane skyrmions exhibit an unidirectional SOT-driven motion regardless of the in-plane orientation of the injected current, which is characterized by the following skyrmion Hall angle
\begin{equation}
\label{SkHangle_in-pl}
\vartheta_{\textrm{SkX}}^{\textrm{in-pl}}=-\left[\tan^{-1}\left(G/\alpha\mathcal{D}\right)+\eta\right].
\end{equation}
Furthermore, their speed along the resultant racetrack is $v=|\partial_{t}\vec{X}|=T_{x}/(G^{2}+\alpha^{2}\mathcal{D}^{2})$, which depends on the $x$-component of the applied current $\vec{j}$; in particular, the speed will be maximum for currents parallel to the easy $x$-axis and zero for currents transverse to it. These findings point towards the design of racetracks for in-plane skyrmions, whose speed could be tuned by adjusting the angle between the charge current and the uniform background magnetization.
We note in passing that the absence of a field-like contribution to the driving force $\vec{T}$ stems from the geometry of this SOT: for an electric current injected along the $x$ direction the induced field-like SOT points along $\vec{e}_{z}\times\vec{j}\propto \hat{e}_{y}$, which tilts the magnetization away from the $z$-axis with no net contribution to the motion of the skyrmion.\cite{Ritzmann2018, Romming2015}

We contrast our results to analogous SOT-driven dynamics of N\'{e}el skyrmions. The rigid hard cutoff ansatz for these solitons takes the form $\bm{m}(\vec{r}\hspace{0.02cm})=(\sin\theta(r)\cos\phi,\sin\theta(r)\sin\phi,\cos\theta(r))^{\top}$, from which we obtain the expressions $T_{k}=\pi^{2}R\tau_{\textrm{DL}}j_{k}/2$, with $k=x,y$. As a result, the skyrmion Hall angle for the spin-orbit dynamics of N\'{e}el skyrmions reads
\begin{equation}
\label{SkHangle_Neel}
\tan\left(\vartheta_{\textrm{SkX}}^{\textrm{N\'{e}el}}+\eta\right)=\frac{\alpha\mathcal{D}\sin(\eta)-G\cos(\eta)}{\alpha\mathcal{D}\cos(\eta)+G\sin(\eta)},
\end{equation}
and therefore explicitly depends on the in-plane orientation of the injected current.\cite{Litzius-NatPhys2017}

We have corroborated our analytical findings by performing micromagnetic simulations using Mumax3 in a thin film geometry with dimensions $256\times256\times1$ nm$^{3}$ in the absence of stray fields. We utilized a charge current pulse of 1 ns width, constant amplitude $j=1\cdot10^{12}$ A/m$^{2}$ and variable in-plane angle $\eta$. The micromagnetic parameters utilized can be found in Table~\ref{Table1}, along with $\alpha=0.3$ and $\tau_{\textrm{DL}}=0.015$ m$^{2}$/A$\cdot$s.\cite{FN5} We observe that the motion of in-plane skyrmions always occurs along the same spatial direction regardless of the orientation of the charge current, meaning that $\vartheta_{\textrm{SkX}}^{\textrm{in-pl}} +\eta$ is a constant, see Eq.~\eqref{SkHangle_in-pl}. Furthermore, as also depicted in the same panel, the speed of in-plane skyrmions along the racetrack is linear with the $x$-component of the current, $j_{x}=j\cos\eta$, which is in agreement with the analytical expression for $v$. Dipolar interactions lead to a reduction of the terminal speed for in-plane skyrmions, as can be deduced from the starred curve in Fig.~\ref{Fig9}(b).

\section{Analytics of the phase diagram}
\label{Sec_Analytics}

In this Section we exploit the minimal model given by Eq.~\eqref{E_functional_in-plane}, complemented with the DM energy functional compatible with the monoclinic system $Cm$, to determine the boundary between the ferromagnetic and helical phases in the space parametrized by the two reduced Dzyaloshinskii coupling constants $g_{1}=\pi D_{xy}^{y}/4\sqrt{AK}$ and $g_{2}=\pi D_{zy}^{y}/4\sqrt{AK}$. First, we cast Eq.~\eqref{E_functional_in-plane} in terms of the dimensionless coordinates $\vec{\rho}=\vec{r}/\sqrt{A/K}$, which reads
\begin{align}
\label{E_functional_in-plane2}
\mathcal{E}_{\textrm{in-pl}}[\bm{m}]=A\int_{\mathcal{S}}&d^{2}\vec{\rho}\,\bigg\{\frac{1}{2}(\vec{\nabla}_{\hspace{-0,07cm}\rho}\bm{m})^{2}+[1-(\bm{m}^{x})^{2}]\\
&+\frac{4g_{1}}{\pi}\mathcal{L}_{\textrm{in-pl},\rho}[\bm{m}]+\frac{4g_{2}}{\pi}\mathcal{L}_{\textrm{N\'{e}el},\rho}[\bm{m}]\bigg\},\nonumber
\end{align}
where the subscript '$\rho$' indicates partial derivation with respect to $\vec{\rho}$. 

Second, it is instructive to study the nature of the helical state arising in monoclinic systems:  for this purpose, we consider the most generic ansatz for a helix in real space, which is described in terms of the normal to the plane of the helix, $\vec{n}=(\cos\phi\sin\theta,\sin\phi\sin\theta,\cos\theta)^{\top}$, and the pitch vector $\vec{q}\neq\vec{0}$, see Fig.~\ref{Fig10}. By casting the magnetization field in a spin frame of reference adjusted to $\vec{n}$, $\bm{m}(\vec{r}\,)=\cos(\vec{q}\cdot\vec{r})\hat{\bm{e}}_{1}+\sin(\vec{q}\cdot\vec{r})\hat{\bm{e}}_{2}+m_{0}\hat{\bm{e}}_{3}$, where $\big\{\hat{\bm{e}}_{1},\hat{\bm{e}}_{2},\hat{\bm{e}}_{3}\equiv\vec{n}\big\}$ is an orthonormal basis in spin space and $m_{0}$ denotes the out-of-plane projection of the magnetization, and by plugging it into Eq.~\eqref{E_functional_in-plane2}, we obtain the following expression for the total (surface) energy density:

\begin{figure}
\begin{center}
\includegraphics[width=1.0\columnwidth]{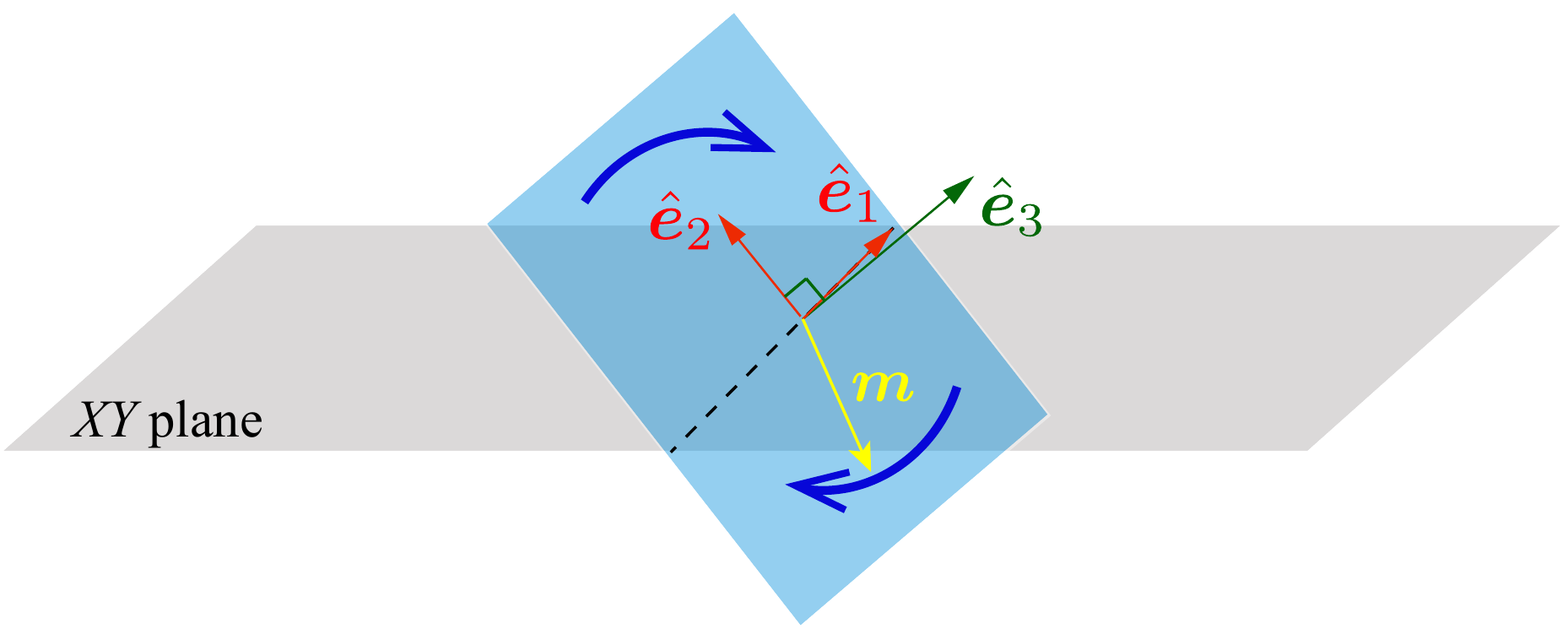}
\caption{Geometry in spin space for the generic helix ansatz of the magnetization field.}
\vspace{-0.5cm} 
\label{Fig10}
\end{center}
\end{figure}

\begin{align}
\label{E_helix}
\varepsilon\big[\bm{m}(\vec{r}\hspace{0.05cm})\big]&=\frac{1}{2}\frac{\vec{q}\hspace{0.07cm}^{2}}{1+m_{0}^{2}}+\frac{4}{\pi}\frac{g_{1}}{1+m_{0}^{2}}\left(q_{y}\cos\theta+q_{x}\sin\theta\sin\phi\right)\nonumber\\
&+\frac{4}{\pi}\frac{g_{2}}{1+m_{0}^{2}}\left(q_{x}\sin\theta\sin\phi-q_{y}\sin\theta\cos\phi\right)\nonumber\\
&+\frac{\frac{1}{2}+m_{0}^{2}}{1+m_{0}^{2}}+\frac{\frac{1}{2}-m_{0}^{2}}{1+m_{0}^{2}}\sin^{2}\theta\cos^{2}\phi.
\end{align}
Extrema of this energy functional with respect to the variables $\{\theta,\phi,\vec{q},m_{0}\}$ are found in Appendix A. Whether any of these solutions corresponds to the ground state of the system can be determined by comparing their energy density~\eqref{E_helix} to that of the uniform magnetic state ($\varepsilon\big[\bm{m}_{u}(\vec{r}\hspace{0.05cm})\big]=0$).

\section{Exactly solvable model at critical coupling}
In this section we consider a minimal exactly solvable model for a two-dimensional ferromagnet hosting stable skyrmions in a uniform magnetization background, the latter pointing along an arbitrary direction within the $xz$ plane, extending therefore the results of Ref.~\onlinecite{Barton-2018} to the case of interfacial-like DM interactions. We note that these skyrmion models are particular instances of the gauge nonlinear $\sigma$ model defined on Riemann surfaces introduced in Ref.~\onlinecite{Schroers-2019}. The resultant family of energy functionals, parametrized by the angle $\alpha$ between the magnetization background and the $z$ axis, includes those stabilizing N\'{e}el skyrmions ($\alpha=0$) and in-plane skyrmions ($\alpha=-\pi/2$). The model can be obtained from Eq. \eqref{E_functional_in-plane} by applying rotations of arbitrary angle $\alpha\in[-\pi,\pi]$ around the $y$ axis in spin space, described by the SO(3) matrix
\begin{equation}
\label{rot_mat}
\hat{R}_{y}[\alpha]=\begin{pmatrix}
\cos\alpha & 0 & \sin\alpha\\
0 & 1 & 0 \\
-\sin\alpha & 0& \cos\alpha
\end{pmatrix}.
\end{equation}
The oriented orthonormal basis associated with the rotated spin frame of reference reads $\hat{e}_{\alpha,k}\equiv\hat{R}_{y}[\alpha]\hat{e}_{k}$, $k=x,y,z$, and the components of the magnetization vector in this frame can be cast as $\bm{m}_{\alpha}^k\equiv\hat{e}_{k}^{\top}\hat{R}_{y}[\alpha]\bm{m}=\hat{e}_{-\alpha,k}\cdot\bm{m}$. The exchange term is invariant under rotations in spin space and the DM energy density becomes
\begin{align}
\label{rot_DMI}
\varepsilon_{\textrm{DM},\alpha}[\bm{m}]&=D  \sum_{i=x,y}\epsilon_{zik}\hat{e}_{-\alpha,k}\cdot(\bm{m}\times\partial_{i}\bm{m}),\\
&=D (\omega_{1}[\bm{m}]+\cos\alpha\,\omega_{2}[\bm{m}]+\sin\alpha\,\omega_{3}[\bm{m}]),\nonumber
\end{align}
in terms of the Lifshitz invariants $\omega_{1}[\bm{m}]=\bm{m}^{z}\partial_{x}\bm{m}^{x}-\bm{m}^{x}\partial_{x}\bm{m}^{z}$, $\omega_{2}[\bm{m}]=\bm{m}^{z}\partial_{y}\bm{m}^{y}-\bm{m}^{y}\partial_{y}\bm{m}^{z}$ and $\omega_{3}[\bm{m}]=\bm{m}^{y}\partial_{y}\bm{m}^{x}-\bm{m}^{x}\partial_{y}\bm{m}^{y}$, where $\epsilon_{zik}$ is the Levi-Civita symbol. We will restrict ourselves to the {\it critical coupling} scenario hereafter, originally introduced in Refs.~\onlinecite{Melcher2014} and~\onlinecite{Barton-2018}, in which the anisotropy barrier is promoted to a 'perfect square' expression in the out-of-plane component of the magnetization via the choice of the appropriate strength for the external magnetic field and Dzyaloshinskii constant, $H=K$ and $D=\sqrt{AK}$  (we disregard stray fields). Consequently, the family of energy functionals for a two-dimensional ferromagnet that we will consider in what follows is
\begin{align}
\label{E_functional_alpha}
\mathcal{E}_{\alpha}[\bm{m}]=A\int_{\mathcal{S}}d^{2}\vec{r}\,&\bigg[\frac{1}{2}(\vec{\nabla}\bm{m})^{2}+\frac{\kappa^{2}}{2}(1-\bm{m}^{z}_{\alpha})^{2}+\\
&\hspace{0.5cm}\kappa \sum_{i=x,y}\epsilon_{zik}\hat{e}_{-\alpha,k}\cdot(\bm{m}\times\partial_{i}\bm{m})\bigg],\nonumber
\end{align}
with $\kappa =D/A$. We can develop a covariant formulation of the model by defining the global Yang-Mills fields $\bm{A}_{j}\equiv \kappa \epsilon_{zjk}\hat{e}_{-\alpha,k}$, $j=x,y$, and then by introducing the covariant (chiral) derivative $\mathcal{D}_{j}\bm{n}\equiv \partial_{j}\bm{n}+\bm{A}_{j}\times\bm{n}=(\partial_{j}-i\bm{A}_{j}\cdot\bm{L})\bm{n}$, where $\bm{n}$ is an arbitrary three-dimensional vector in spin space and $[\hat{L}_{\alpha}]_{\beta\gamma}\equiv -i\epsilon_{\alpha\beta\gamma}$ are the generators of the SO(3) group. The family of energy functionals of Eq.~\eqref{E_functional_alpha} can be cast as (see Appendix B):
\begin{align}
\label{E_functional_alpha_cov}
\mathcal{E}_{\alpha}[\bm{m}]=&\frac{A}{2}\int_{\mathcal{S}}d^{2}\vec{r}\hspace{0.1cm}\big(\mathcal{D}_{x}\bm{m}+\bm{m}\times \mathcal{D}_{y}\bm{m}\big)^{2}\\
&+4\pi A(Q[\bm{m}]+\Omega[\bm{m}]),\nonumber
\end{align}
where we have introduced the functionals
\begin{eqnarray}
\label{top_charge}
Q[\bm{m}]&\equiv&\frac{1}{4\pi}\int_{\mathcal{S}}d^{2}\vec{r}\hspace{0.1cm}\bm{m}\cdot(\partial_{x}\bm{m}\times\partial_{y}\bm{m}),\\
\Omega[\bm{m}]&\equiv&
\frac{\kappa}{4\pi}\int_{\mathcal{S}}d^{2}\vec{r}\hspace{0.1cm} \nabla \cdot \bm{m}_{\alpha}=\frac{\kappa}{4\pi}\int_{\partial \mathcal{S}}\big(\bm{m}_{\alpha}^{x}dy-\bm{m}_{\alpha}^{y}dx\big),\nonumber
\label{boundary_term}
\end{eqnarray}
representing the topological and 'total magnetic' charges of the spin texture, respectively. The expression in Eq.~\eqref{E_functional_alpha_cov} is particularly useful, as the first term is always positive definite because of the quadratic expression. Therefore, any state with a fixed topological charge $Q$ and magnetic charge $\Omega$ has the energy $\mathcal{E}_{\alpha0}=4\pi A(Q+\Omega)$ as a lower bound, which is equivalent to solving the Bogomol'nyi equations $\mathcal{D}_{x}\bm{m}+\bm{m}\times \mathcal{D}_{y}\bm{m}=\bm{0}$, which can be recast as (see Appendix C):
\begin{equation}
\label{Bogomol'nyi_eqs}
\bm{m}\times\Delta\bm{m}=2\kappa(\bm{m}_{\alpha}^{y}\partial_{x}\bm{m}-\bm{m}_{\alpha}^{x}\partial_{y}\bm{m})+\kappa^{2}(1-\bm{m}_{\alpha}^{z})\hat{e}_{-\alpha,z}\times\bm{m},
\end{equation}
where $\Delta$ denotes the Laplacian operator. Note that spatially localized magnetic textures (namely, those extending to continuous maps $S^{2}\rightarrow S^{2}$) have zero total magnetic charge, $\Omega[\bm{m}]=0$, and, therefore, their total energy equals the Bogomol'nyi bound $\mathcal{E}_{\alpha}=4\pi AQ$.

\begin{figure}
\begin{center}
\includegraphics[width=1.0\columnwidth]{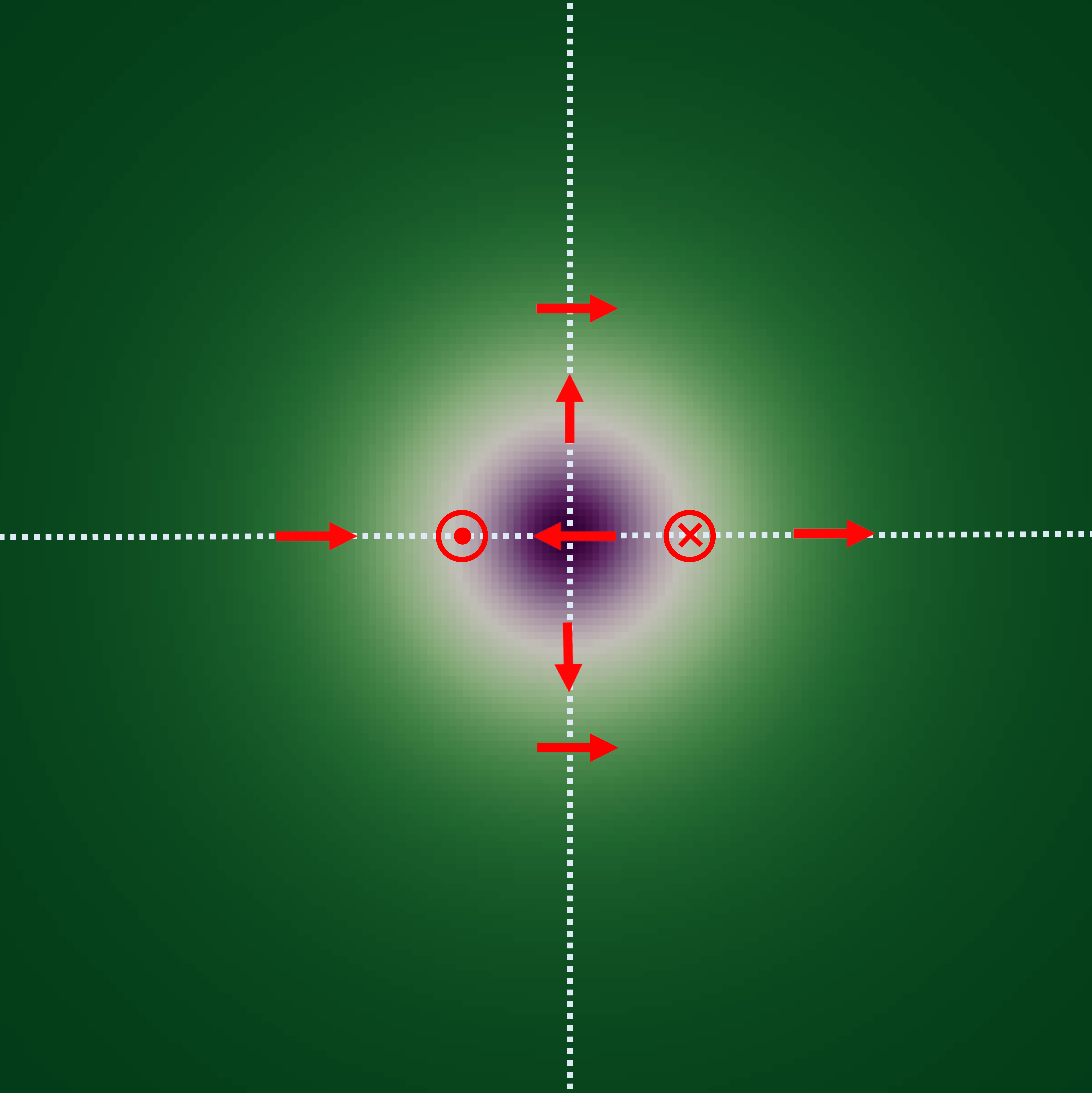}
\caption{Spatial dependence of the magnetization for an isolated in-plane skyrmion at critical coupling and disregarding demagnetization effects. The ratio of the Dzyaloshinskii constant to the spin stiffness has been taken to be $\kappa=1/2$. The color code indicates the $\bm{m}^{x}$ component of the magnetization.}
\vspace{-0.5cm} 
\label{Fig11}
\end{center}
\end{figure}

Analysis of the Bogomol'nyi equations \eqref{Bogomol'nyi_eqs} simplifies if we consider the stereographic projection of the magnetization vector (in the rotated spin frame of reference) onto the complex plane:
\begin{equation}
\label{stereo_proj}
\xi=\frac{\bm{m}_{\alpha}^{x}+i\bm{m}_{\alpha}^{y}}{1+\bm{m}_{\alpha}^{z}},
\end{equation}
where $\bm{m}:\mathbb{C}\rightarrow S^{2}$ is now defined over the field of complex numbers. The family of energy functionals \eqref{E_functional_alpha} can be cast in terms of this complex representation of the order parameter as (see Appendix D):
\begin{align}
\label{E_functional_alpha_stereo}
\mathcal{E}_{\alpha}[\xi]
=&8A\int_{\mathcal{S}}d^{2}\vec{r}\frac{\big(\partial_{\zeta}\bar{\xi}+\frac{\kappa}{2}\bar{\xi}\hspace{0.03cm}^{2}\big)\big(\partial_{\bar{\zeta}}\xi+\frac{\kappa}{2}\xi^{2}\big)}{(1+|\xi|^{2})^{2}}\\
&+4\pi A(Q[\xi]+\Omega[\xi]),\nonumber
\end{align}
where $\zeta=x+iy$ and the topological charge and the 'total magnetic' charge take the form
\begin{align}
Q[\xi]&=\frac{i}{2\pi}\int_{\mathcal{S}}d^{2}\vec{r}\,\frac{\partial_{x}\xi\partial_{y}\bar{\xi}-\partial_{y}\xi\partial_{x}\bar{\xi}}{(1+|\xi|^{2})^{2}},\\
\Omega[\xi]&=\frac{\kappa}{\pi}\int_{\mathcal{S}}d^{2}\vec{r}\,\frac{\textrm{Re}\big[\partial_{\zeta}\xi-\xi^{2}\partial_{\zeta}\bar{\xi}\hspace{0.05cm}\big]}{(1+|\xi|^{2})^{2}}.\nonumber
\end{align}
 The Bogomol'nyi equations then become
\begin{equation}
\label{Bogomol'nyi_eqs_complex}
\partial_{\bar{\zeta}}\xi=-\frac{\kappa}{2}\xi^{2}\stackrel{w=1/\xi}{\Longrightarrow}\partial_{\bar{\zeta}}w=\frac{\kappa}{2},
\end{equation}
which can be solved exactly as $w=\frac{\kappa}{2}\bar{\zeta}+f(\zeta)$, where $f:\mathbb{C}\rightarrow\mathbb{C}$ denotes any holomorphic function. 

In-plane skyrmions, also known as magnetic bimerons, are metastable soliton solutions of the model \eqref{E_functional_alpha} for $\alpha=-\pi/2$. The analytic expression for an isolated in-plane skyrmion is given by the solution $\xi=2/\kappa\bar{\zeta}$ of the Bogomol'nyi equations. The corresponding magnetization vector reads (see Appendix D)
\begin{equation}
\label{in-plane_sky}
\bm{m}_{x}=\frac{\kappa^{2}|\zeta|^{2}-4}{\kappa^{2}|\zeta|^{2}+4},\hspace{0.2cm}\bm{m}_{z}-i\bm{m}_{y}=-\frac{4\kappa\zeta}{\kappa^{2}|\zeta|^{2}+4},
\end{equation}
and Figure \ref{Fig11} depicts the spatial profile of an in-plane skyrmion for the value $\kappa=1/2$.

\section{Conclusions}

Even though Bloch-like and N\'{e}el-like configurations represent the archetypical skyrmions found in conventional chiral magnets and magnetic bilayers subjected to (strong) spin-orbit coupling at the interface, infinite many other families of skyrmion configurations can, in principle, be smoothly generated from the former via proper rotations in spin space while preserving their topological robustness, e.g., in-plane skyrmions. Remarkably, as the findings of our study indicate, the behavior of these latter families is qualitatively similar to that of N\'{e}el/Bloch skyrmions, although some features remain inherently distinct as a result of the additional rotational symmetry breaking (along the $z$ axis).

In this paper, we have shown first that the monoclinic system $Cm$ is the only one compatible with the symmetries of in-plane skyrmions and proposed material candidates, such as FeLa$_{3}$S$_{6}$ and Rb$_{6}$Fe$_{2}$O$_{5}$, to host this class of topological solitons. We have also predicted the (meta)stability of in-plane skyrmions in a wide area of the phase diagram, the latter being parametrized by the two reduced Dzyaloshinkii coupling constants describing the monoclinic chiral interaction. Remarkably, the presence of an interfacial DM stabilizer enhances the stability of in-plane skyrmions regardless of its strength. Second, we have proved via micromagnetic simulations that the effect of magnetostatic interactions on the stability of skyrmions is substantially stronger for in-plane configurations than for N\'{e}el-like ones. For instance, in the presence of stray fields, in-plane skyrmions exhibit smaller sizes than those of the N\'{e}el type for the same set of micromagnetic parameters; as a result, in-plane skyrmions appear to be more appealing from the technological standpoint than their N\'{e}el counterparts, since they favor the development of skyrmion devices with higher (spatial) density. In this regard, dipolar fields lying in the plane of the magnet allows for the vertical stacking of in-plane skyrmion layers, which represents a major challenge for PMA magnets hosting N\'{e}el skyrmions.\cite{Gobel-PRB2019} Furthermore, the skyrmion radius is ill-defined for in-plane configurations since dipolar interactions significantly alter their shape with respect to the axially-symmetric one.

Third, we have corroborated that the methods for skyrmion production based on DW blowing through a geometric constriction and shedding from a magnetic impurity, which were originally devised in the N\'{e}el scenario, also work for the in-plane case. In particular, for the shedding method, the critical current is of the same order of magnitude $(\sim10^{12}$ A/m$^{2})$ but larger than that of N\'{e}el-like configurations  due to action of stray fields, and antiskyrmions are rapidly annihilated as a result of the in-plane DM interaction. For the DW blowing method, in-plane skyrmions are produced for thin enough (as compared to the sample width) constrictions and magnetostatic interactions do not qualitatively affect the creation process. Fourth, SOT-driven dynamics for in-plane skyrmions differ from those for N\'{e}el-like configurations: strikingly, field-like torques do not contribute to the dynamics of in-plane skyrmions, which, in turn, yields the unidirectional character of their motion. As a result, materials such as FeLa$_{3}$S$_{6}$ and Rb$_{6}$Fe$_{2}$O$_{5}$, which can potentially host these solitons, are optimal platforms for the skyrmion racetrack principle to be exploited. On the other hand, spin-transfer dynamics are similar for in-plane and N\'{e}el skyrmions. Furthermore, dipolar interactions yield a reduction of the (terminal) speed for both spin-transfer and SOT-driven dynamics. Table~\ref{Table2} compares different features of skyrmions arising in both in-plane and N\'{e}el scenarios, providing a brief summary of the findings just discussed.

\begin{table}[ht!]
\begin{center}
\label{Tab1}
  \begin{tabular}{ c | c | c |}
  \cline{2-3}
  & in-plane & N\'{e}el \\
     \hline
   \multicolumn{1}{|c|}{\textrm{Crystal symmetries}}& $Cm$ & $C_{nv}$ \\ \hline
     \multicolumn{1}{|c|}{\textrm{Phase diagram}}& 2D & 1D \\ \hline
    \multicolumn{1}{|c|}{\textrm{Production via blowing}}& Yes & Yes \\ \hline
     \multicolumn{1}{|c|}{\textrm{Production via shedding}}& Yes & Yes \\ \hline
      \multicolumn{1}{|c|}{\textrm{Antidamping-like torque contrib.}}& Yes & Yes \\ \hline
       \multicolumn{1}{|c|}{\textrm{Field-like torque contrib.}}& No & No \\ \hline
      \multicolumn{1}{|c|}{\textrm{Racetrack principle}}& Yes & No \\ \hline
 \end{tabular}
  \caption{Comparison table between in-plane and N\'{e}el skyrmions.}
  \label{Table2}
\end{center}
\end{table}

Fourth, we have built an exactly solvable model for a continuous family of skyrmion configurations, including the in-plane ones, parametrized by the rotation angle along the $y$ axis in spin space. We have derived the Bogomol'nyi equations describing the spatial structure of these solitons and solved them analytically. We have shown that these solutions agree well with in-plane skyrmions created via micromagnetic simulations. Our ansatz may be useful, e.g., to estimate transport coefficients or derive equations of motion within a collective variable approach in future research.

Finally, in this work we have focused on the stability and dynamics of individual skyrmions at zero temperature. On the one hand, skyrmion-skyrmion and skyrmion-edge repulsions play a significant role in the experimental realization of skyrmion-based racetrack memories. The effect of repulsive forces in N\'{e}el-skyrmion-based racetracks has been previously investigated in Ref.~\onlinecite{Zhang-SciRep2015}, which suggests that skyrmion-skyrmion repulsion can be made very small/negligible by tuning the anisotropy and the external magnetic field, so that the spacing between skyrmions within the racetrack will remain approximately constant. Even though we could {\it a priori} expect a similar tunability for in-plane skyrmions, the effect of dipolar fields and repulsive forces is not obvious in this case and can be investigated thoroughly via micromagnetic simulations. On the other hand, we expect for in-plane skyrmions a similar robustness against thermal fluctuations to that of the N\'{e}el type. In the latter case, this thermal robustness has been shown on a racetrack in Refs.~\onlinecite{Muller-NJP2017} and~\onlinecite{Suess-SciRep2019}. A detailed study of the effect of repulsive interactions and thermal fluctuations on the properties of in-plane skyrmions goes beyond the scope of this work and will be carried out in future works.

\acknowledgements

The authors thank L.~\v{S}mejkal, B.~F.~McKeever  and D.~R.~Rodrigues for useful discussions. This work has been supported by the Alexander von Humboldt Foundation, the Transregional Collaborative Research Center (SFB/TRR) 173 SPIN+X, the Emmy Noether: German Research Foundation (DFG) Project No. EV 196/2-1, the skyrmionics SPP: EV196/5-1, SI1720/4-1 and CHIME Project No.~SI1720/2-1.

R.Z. and V.K.B. contributed equally to this work. 
\vspace{0.5cm}


\appendix

\section{Derivation of Eq.~(13)}

Extremalization of the energy density functional~\eqref{E_helix} yields the identities
\begin{align}
\label{q_pitch}
q_{x}&=-\frac{4}{\pi}(g_{1}+g_{2})\sin\theta\sin\phi,\\
q_{y}&=\frac{4}{\pi}\left(g_{2}\sin\theta\cos\phi-g_{1}\cos\theta\right),
\end{align}
for the pitch vector and the value $m_{0}\equiv0$ for the out-of-plane magnetization. Furthermore, the polar angle $\theta$ satisfies either the condition $\theta=0,\pi$ or the identity 
\begin{align}
\label{eq_helix3}
\delta g_{1}g_{2}\left[2g_{1}g_{2}-\frac{16}{\pi^{2}}\delta^{2}\right]&\cos2\theta\cot\theta-\delta^{3}\cot\theta\\
&+g_{1}^{2}g_{2}^{3}(2g_{1}+g_{2})\sin2\theta=0,\nonumber
\end{align}
with $\delta\equiv g_{1}^{2}+2g_{1}g_{2}+\pi^{2}/16$. In the former case, the resultant magnetic texture reads
\begin{equation}
\label{ansatz1}
\bm{m}(\vec{r}\hspace{0.05cm})=\pm\left(\cos[4g_{1}y/\pi],-\sin[4g_{1}y/\pi],0\right)^{\top}
\end{equation}
and its total energy density becomes $\varepsilon=\frac{1}{2}\left[1-\frac{16g_{1}^{2}}{\pi^{2}}\right]$. In the latter case, the magnetization field is parametrized by the polar and azimuthal angles resulting from the solution of Eqs.~\eqref{eq_helix3} and 
\begin{equation}
\label{eq_helix4}
\cos\phi=-\frac{g_{1}^{2}+2g_{1}g_{2}+\pi^{2}/16}{g_{1}g_{2}}\cot\theta,
\end{equation}
from which the corresponding total energy density can be calculated by means of Eq.~\eqref{E_helix}.

\section{Derivation of Eq.~(17)}

The covariant expression \eqref{E_functional_alpha_cov} can be obtained, as discussed in Refs.~\onlinecite{Barton-2018} and~\onlinecite{Schroers-2019}, by noticing that
\begin{align}
\label{eq_aux_1}
(\mathcal{D}_{x}\bm{m})^{2}+(\mathcal{D}_{y}&\bm{m})^{2}=(\vec{\nabla}\bm{m})^{2}+\kappa^{2}\big[1+(\bm{m}^{z}_{\alpha})^{2}\big]\\
&+2\kappa\sum_{i=x,y}\epsilon_{zik}\hat{e}_{-\alpha,k}\cdot(\bm{m}\times\partial_{i}\bm{m}),\nonumber
\end{align}
and that
\begin{align}
\label{eq_aux_2}
\frac{1}{2}&\big(\mathcal{D}_{x}\bm{m}+\bm{m}\times \mathcal{D}_{y}\bm{m}\big)^{2}=\frac{1}{2}(\mathcal{D}_{x}\bm{m})^{2}+\frac{1}{2}(\mathcal{D}_{y}\bm{m})^{2}\\
&-\bm{m}\cdot(\partial_{x}\bm{m}\times\partial_{y}\bm{m})-\kappa(\partial_{x}\bm{m}_{\alpha}^{x}+\partial_{y}\bm{m}_{\alpha}^{y})-\kappa^{2}\bm{m}_{\alpha}^{z}.\nonumber
\end{align}
The latter identity can be derived by noting that the Yang-Mills fields $\bm{A}_{x,y}$ engender the antisymmetric Faraday tensor $\bm{F}_{xy}=\partial_{x}\bm{A}_{y}-\partial_{y}\bm{A}_{x}+\bm{A}_{x}\times\bm{A}_{y}=\kappa^{2}\hat{e}_{-\alpha,z}$ and that i) $\bm{m}\cdot\bm{A}_{x}=\kappa \,\bm{m}_{\alpha}^{y}$ and $\bm{m}\cdot\bm{A}_{y}=-\kappa\,\bm{m}_{\alpha}^{x}$, ii) $(\mathcal{D}_{x}\bm{m}+\bm{m}\times \mathcal{D}_{y}\bm{m})^{2}=(\mathcal{D}_{x}\bm{m})^{2}+(\mathcal{D}_{y}\bm{m})^{2}-2\bm{m}\cdot(\mathcal{D}_{x}\bm{m}\times \mathcal{D}_{y}\bm{m})$ and iii) the identity
\begin{align}
\bm{m}\cdot(\mathcal{D}_{x}\bm{m}\times \mathcal{D}_{y}\bm{m})&=\bm{m}\cdot(\partial_{x}\bm{m}\times \partial_{y}\bm{m})+\partial_{y}(\bm{m}\cdot\bm{A}_{x})-\nonumber\\
&\hspace{0.7cm}\partial_{x}(\bm{m}\cdot\bm{A}_{y})+\bm{m}\cdot\bm{F}_{xy}
\end{align}
also holds.

\section{Derivation of Bogomol'nyi equations}

By making explicit the expression for the covariant derivatives, the Bogomol'nyi equations read
\begin{align}
\label{Bogomol'nyi_eqs_1}
\partial_{x}\bm{m}&=-\bm{m}\times\partial_{y}\bm{m}+\kappa[\bm{m}\times (\hat{e}_{-\alpha,x}\times\bm{m})-\hat{e}_{-\alpha,y}\times\bm{m}],\\
\label{Bogomol'nyi_eqs_2}
\partial_{y}\bm{m}&=\bm{m}\times\partial_{x}\bm{m}+\kappa[\bm{m}\times(\hat{e}_{-\alpha,y}\times\bm{m})+\hat{e}_{-\alpha,x}\times\bm{m}],
\end{align}
where the second equation is obtained by applying $\bm{m}\times$ to the first one. The cross product $\bm{m}\times\Delta\bm{m}=\bm{m}\times[\partial_{x}^{2}\bm{m}+\partial_{y}^{2}\bm{m}]$ can therefore be cast as
\begin{align}
\label{Bogomol'nyi_Laplacian}
\bm{m}&\times\Delta \bm{m}=2\kappa(\bm{m}_{\alpha}^{y}\partial_{x}\bm{m}-\bm{m}_{\alpha}^{x}\partial_{y}\bm{m})+\kappa^{2}(\bm{m}_{\alpha}^{x}\hat{e}_{-\alpha,y}\nonumber\\
&-\bm{m}_{\alpha}^{y}\hat{e}_{-\alpha,x}+\bm{m}_{\alpha}^{x}\hat{e}_{-\alpha,x}\times\bm{m}+\bm{m}_{\alpha}^{y}\hat{e}_{-\alpha,y}\times\bm{m}),
\end{align}
and, with account of the identity $\bm{m}_{\alpha}^{x}\hat{e}_{-\alpha,x}\times\bm{m}+\bm{m}_{\alpha}^{y}\hat{e}_{-\alpha,y}\times\bm{m}+\bm{m}_{\alpha}^{x}\hat{e}_{-\alpha,y}-\bm{m}_{\alpha}^{y}\hat{e}_{-\alpha,x}=(1-\bm{m}_{\alpha}^{z})\hat{e}_{-\alpha,z}\times\bm{m}$, we finally obtain Eq.~\eqref{Bogomol'nyi_eqs}.

\section{Details of the stereographic projection}

We start this Appendix by noting that the inverse transformation to Eq.~\eqref{stereo_proj} reads
\begin{equation}
\label{inverse_stereo_proj}
\bm{m}_{\alpha}^{x}+i\bm{m}_{\alpha}^{y}=\frac{2\xi}{1+|\xi|^{2}},\hspace{0.2cm} \bm{m}_{\alpha}^{z}=\frac{1-|\xi|^{2}}{1+|\xi|^{2}}.
\end{equation}
Then, the expression \eqref{E_functional_alpha_stereo} for energy functional is derived with account of the identity
\begin{equation}
\hspace{-0.08cm}\frac{\mathcal{E}_{\alpha}[\xi]}{2A}=\hspace{-0.05cm}\int_{\mathcal{S}} d^{2}\vec{r}\,\frac{|\vec{\nabla}\xi|^{2}\hspace{-0.05cm}+\hspace{-0.05cm}2\kappa\textrm{Re}[\partial_{\zeta}\xi\hspace{-0.05cm}+\hspace{-0.05cm}\xi^{2}\partial_{\zeta}\bar{\xi}\hspace{0.05cm}]+\kappa^{2}|\xi|^{4}}{(1+|\xi|^{2})^{2}},
\end{equation}
and the following mathematical relations:
\begin{widetext}
\begin{align}
(1-\bm{m}_{\alpha}^{z})^{2}&=\frac{4|\xi|^{4}}{(1+|\xi|^{2})^{2}},\\
|\partial_{k}\bm{m}|^{2}&=|\partial_{k}(\bm{m}_{\alpha}^{x}+i\bm{m}_{\alpha}^{y})|^{2}+(\partial_{k}\bm{m}_{\alpha}^{z})^{2}=\frac{4|\partial_{k}\xi|^{2}}{(1+|\xi|^{2})^{2}},\\
\partial_{x}\bm{m}_{\alpha}^{x}+\partial_{y}\bm{m}_{\alpha}^{y}&=2\textrm{Re}\big[\partial_{\zeta}(\bm{m}_{\alpha}^{x}+i\bm{m}_{\alpha}^{y})\big]=\frac{4}{(1+|\xi|^{2})^{2}}\textrm{Re}\big[\partial_{\zeta}\xi-\xi^{2}\partial_{\zeta}\bar{\xi}\hspace{0.05cm}\big],\\
\sum_{i=x,y}(\hat{e}_{z}\times\hat{e}_{i})\cdot(\bm{m}_{\alpha}\times\partial_{i}\bm{m}_{\alpha})&=\textrm{Re}\big[\bm{m}_{\alpha}^{z}\partial_{\zeta}(\bm{m}_{\alpha}^{x}+i\bm{m}_{\alpha}^{y})-(\bm{m}_{\alpha}^{x}+i\bm{m}_{\alpha}^{y})\partial_{\zeta}\bm{m}_{\alpha}^{z}\big]=
\frac{4}{(1+|\xi|^{2})^{2}}\textrm{Re}\big[\partial_{\zeta}\xi+\xi^{2}\partial_{\zeta}\bar{\xi}\hspace{0.04cm}\big],\\
\bm{m}\cdot(\partial_{x}\bm{m}\times\partial_{y}\bm{m})&=2i\frac{\partial_{x}\xi\partial_{y}\bar{\xi}-\partial_{y}\xi\partial_{x}\bar{\xi}}{(1+|\xi|^{2})^{2}},
\end{align}
\end{widetext}
where $\partial_{\zeta}\equiv\frac{1}{2}(\partial_{x}-i\partial_{y})$.


\begin{thebibliography}{99}

\bibitem{Belavin-JETP1975} A. A. Belavin and A. M. Polyakov, JETP Lett. {\bf 22}, 245 (1975).

\bibitem{Bogdanov-PRB2002} A. N. Bogdanov, U. K. R\"ossler, M. Wolf and K.-H. M\"uller, Phys. Rev. B {\bf 66}, 214410 (2002).

\bibitem{Banerjee-PRX2014} S. Banerjee, J. Rowland, O. Erten and M. Randeria, Phys. Rev. X {\bf 4}, 031045 (2014).

\bibitem{Mulbauer-Sci2009} S. M\"{u}hlbauer, B. Binz, F. Jonietz, C. Pfleiderer, A. Rosch, A. Neubauer, R. Georgii and P. B\"{u}ni, Science {\bf 323}, 915 (2009).

\bibitem{Yu-Nat2010} X. Z. Yu, Y. Onose, N. Kanazawa, J. H. Park, J. H. Han, Y. Matsui, N. Nagaosa and Y. Tokura, Nature (London) {\bf 465}, 901 (2010).

\bibitem{Seki-Sci2012} S. Seki, X. Z. Yu, S. Ishiwata and Y. Tokura, Science {\bf 336}, 198 (2012).

\bibitem{Romming-Sci2013} N. Romming, C. Hanneken, M. Menzel, J. E. Bickel, B. Wolter, K. von Bergmann, A. Kubetzka and R. Wiesendanger, Science {\bf 341}, 636 (2013).

\bibitem{Jiang-Sci2015} W. Jiang, P. Upadhyaya, W. Zhang, G. Yu, M. B. Jungfleisch, F. Y. Fradin, J. E. Pearson, Y. Tserkovnyak, K. L. Wang, O. Heinonen, S. G. E. te Velthuis and A. Hoffmann, Science {\bf 349}, 283 (2015).

\bibitem{Dzyaloshinskii-JETP1957} I. E. Dzyaloshinskii, Sov. Phys. JETP {\bf 5}, 1259 (1957).

\bibitem{Moriya-PR1960} T. Moriya, Phys. Rev. {\bf 120}, 91 (1960).

\bibitem{Dzyaloshinskii-JETP1964} I. E. Dzyaloshinskii, Sov. Phys. JETP {\bf 19}, 960 (1964).

\bibitem{Fert-PRL1980} A. Fert and P. M. Levy, Phys. Rev. Lett. {\bf 44}, 1538 (1980)

\bibitem{Leonov-NatComms2015} A. Leonov and M. Mostovoy, Nat. Commun. {\bf 6}, 8275 (2015).

\bibitem{Lin-PRB2016} S.-Z. Lin and S. Hayami, Phys. Rev. B {\bf 93}, 064430 (2016).

\bibitem{Bezvershenko-PRB2019} A. V. Bezvershenko, A. K. Kolezhuk and B. A. Ivanov, Phys. Rev. B {\bf 97}, 054408 (2019).

\bibitem{Parkin-Science2008} S. S. P. Parkin, M. Hayashi, and L. Thomas, Science {\bf 320}, 190 (2008).

\bibitem{Nagaosa-NatNano2013} N. Nagaosa and Y. Tokura, Nat. Nanotechnol. {\bf 8}, 899 (2013), and references therein.

\bibitem{Zhou-NatComm2014} Y. Zhou and M. Ezawa, Nat. Commun. {\bf 5}, 4652 (2014).

\bibitem{Zhang-SciRep2015} X. Zhang, M. Ezawa and Y. Zhou, Sci. Rep. {\bf 5}, 9400 (2015).

\bibitem{Fert-NatRevMat2017} A. Fert, N. Reyren, and V. Cros, Nat. Rev. Mat. {\bf 2}, 17031 (2017).

\bibitem{Huang-Nanotech2017} Y. Huang, W. Kang, X. Zhang, Y. Zhou, and W. Zhao, Nanotechnology {\bf 28}, 08LT02 (2017).

\bibitem{Li-Nanotech2017} S. Li, W. Kang, Y. Huang, X. Zhang, Y. Zhou, and W. Zhao, Nanotechnology {\bf 28}, 31LT01 (2017).

\bibitem{Prychynenko-PRAppl2018} D. Prychynenko, M. Sitte, K. Litzius, B. Kr\"{u}ger, G. Bourianoff, M. Kl\"{a}ui, J. Sinova, and K. Everschor-Sitte, Phys. Rev. Appl. {\bf 9}, 014034 (2018).

\bibitem{Azam-JApplPhys2018} Md. A. Azam, D. Bhattacharya, D. Querlioz, and J. Atulasimha, J. Appl. Phys. {\bf 124}, 152122 (2018).

\bibitem{Bourianoff-AIPAdv2018} G. Bourianoff, D. Pinna, M. Sitte, and K. Everschor-Sitte, AIP Adv. {\bf 8}, 055602 (2018).

\bibitem{Pinna2019} D. Pinna, G. Bourianoff, K. Everschor-Sitte, arXiv:1811.12623(2018).

\bibitem{Zazvorka2019} J. Z\'azvorka, F. Jakobs, D. Heinze, N. Keil, S. Kromin, S. Jaiswal, K. Litzius, G. Jakob, P. Virnau, D. Pinna, K. Everschor-Sitte, L. R\'ozsa, A. Donges, U. Nowak and M. Kl\'aui, Nat. Nanotechnol. {\bf 14}, 658 (2019).

\bibitem{Pollath-PRL2017} S. P\"{o}llath, J. Wild, L. Heinen, T. N. G. Meier, M. Kronseder, L. Tutsch, A. Bauer, H. Berger, C. Pfleiderer, J. Zweck, A. Rosch, and C. H. Back, Phys. Rev. Lett. {\bf 118}, 207205 (2017). 

\bibitem{Lin-PRB2013} S.-Z. Lin, C. Reichhardt, C. D. Batista and A. Saxena, Phys. Rev. B {\bf 87}, 214419 (2013).

\bibitem{Jonietz-Science2010} F. Jonietz, S. M\"{u}hlbauer, C. Pfleiderer, A. Neubauer, W. M\"{u}nzer, A. Bauer, T. Adams, R. Georgii, P. B\"{o}ni, R. A. Duine, K. Everschor, M. Garst and A. Rosch, Science {\bf 330}, 1648 (2010)

\bibitem{Schulz-NatPhys2012} T. Schulz, R. Ritz, A. Bauer, M. Halder, M. Wagner, C. Franz, C. Pfleiderer, K. Everschor, M. Garst and A. Rosch, Nat. Phys. {\bf 8}, 301 (2012).

\bibitem{Fert-NatNano2013} A. Fert, V. Cros, and J. Sampaio, Nat. Nanotechnol. {\bf 8}, 152 (2013).

\bibitem{Sitte-PRB2016} M. Sitte, K. Everschor-Sitte, T. Valet, D. R. Rodrigues, J. Sinova and Ar. Abanov, Phys. Rev. B {\bf 94}, 064422 (2016).

\bibitem{Everschor-Sitte-NJP2017} K. Everschor-Sitte, M. Sitte, T. Valet, Ar. Abanov and J. Sinova, New J. Phys. {\bf 19}, 092001 (2017).

\bibitem{Stier-PRL2017} M. Stier, W. H\"{a}usler, T. Posske, G. Gurski and M. Thorwart, Phys. Rev. Lett. {\bf 118}, 267203 (2017).

\bibitem{Buttner-NNano2017} F. B\"{u}ttner, I. Lemesh, M. Schneider, B. Pfau, C. M. G\"{u}nther, P. Hessing, J. Geilhufe, L. Caretta, D. Engel, B. Kr\"{u}ger, J. Viefhaus, S. Eisebitt and G. S. D. Beach, Nature Nanotechnology {\bf 12}, 1040 (2017).

\bibitem{Sampaio-NatNano2013} J. Sampaio, V. Cros, S. Rohart, A. Thiaville and A. Fert, Nat. Nanotechnol. {\bf 8} 839 (2013).

\bibitem{Iwasaki-NatComms2013} J. Iwasaki, M. Mochizuki, and N. Nagaosa, Nat. Comms. {\bf 4}, 1463 (2013)

\bibitem{Rohart-PRB2016} S. Rohart, J. Miltat, and A. Thiaville, Phys. Rev. B {\bf 93}, 214412 (2016).

\bibitem{Jiang-NatPhys2017} W. Jiang, X. Zhang, G. Yu, W. Zhang, X. Wang, M. B. Jungfleisch, J. E. Pearson, X. Cheng, O. Heinonen, K. L. Wang, Y. Zhou, A. Hoffmann and S. G. E. te Velthuis, Nat. Phys. {\bf 13}, 162 (2017). 


\bibitem{Romming2015} N. Romming, A. Kubetzka, C. Hanneken, K. von Bergmann and R. Wiesendanger, Phys. Rev. Lett. {\bf 114}, 177203 (2015).

\bibitem{Ritzmann2018} U. Ritzmann, S. von Malottki, J.-V. Kim, S. Heinze, J. Sinova, and B. Dup\'e, Nat. Electron. {\bf 1}, 451 (2018).



\bibitem{Litzius-NatPhys2017} K. Litzius, I. Lemesh, B. Kr\"{u}ger, P. Bassirian, L. Caretta, K. Richter, F. B\"{u}ttner, K. Sato, O. A. Tretiakov, J. F\"{o}rster, R. M. Reeve, M. Weigand, I. Bykova, H. Stoll, G. Schütz, G. S. D. Beach and M. Kl\"{a}ui, Nat. Phys. {\bf 13}, 170 (2017).

\bibitem{Bruno-PRL2004} P. Bruno, V. K. Dugaev, and M. Taillefumier, Phys. Rev. Lett. {\bf 93}, 096806 (2004).

\bibitem{Neubauer-PRL2009} A. Neubauer, C. Pfleiderer, B. Binz, A. Rosch, R. Ritz, P. G. Niklowitz, and P. B\"{o}ni, Phys. Rev. Lett. {\bf 102}, 186602 (2009).

\bibitem{Yokouchi-JPSJ2015} T. Yokouchi, N. Kanazawa, A. Tsukazaki, Y. Kozuka, A. K., Y. Taguchi, M. Kawasaki, M. Ichikawa, F. Kagawa, and Y. Tokura, J. Phys. Soc. Jpn. {\bf 84}, 104708 (2015).

\bibitem{Meynell-PRB2017} S. A. Meynell, M. N. Wilson, K. L. Krycka, B. J. Kirby, H. Fritzsche, and T. L. Monchesky, Phys. Rev. B {\bf 96}, 054402 (2017).

\bibitem{Kharkov-PRL2017} Y. A. Kharkov, O. P. Sushkov, and M. Mostovoy, Phys. Rev. Lett. {\bf 119}, 207201 (2017).


\bibitem{FN0}  These topological textures, being the in-plane analogues of conventional N\'{e}el skyrmions, are intrinsically different to the radial vortices reported in Ref.~\onlinecite{Siracusano-PRL2016}, the latter being skyrmion-like magnetic configurations with noninteger topological charge due to the DM-induced tilting of the magnetization at the boundaries of the sample.

\bibitem{Siracusano-PRL2016} G. Siracusano, R. Tomasello, A. Giordano, V. Puliafito, B. Azzerboni, O. Ozatay, M. Carpentieri, and G. Finocchio, Phys. Rev. Lett. {\bf 117}, 087204 (2016). 

\bibitem{Gobel-PRB2019} B. G\"{o}bel, A. Mook, J. Henk, I. Mertig, and O. A. Tretiakov, Phys. Rev. B {\bf 99}, 060407(R) (2019). 

\bibitem{Shen-2019} L. Shen, J. Xia, X. Zhang, M. Ezawa, O. A. Tretiakov, X. Liu, G. Zhao, Y. Zhou, arXiv:1905.09007 (2019).

\bibitem{Heo-SciRep2016} C. Heo, N. S. Kiselev, A. K. Nandy, S. Bl\"{u}gel, and T. Rasing, Sci. Rep. {\bf 6}, 27146 (2016).

\bibitem{Barton-2018} B. Barton-Singer, C. Ross and B. J. Schroers, arXiv:\\1812.07268 (2018).

\bibitem{Schroers-2019} B. J. Schroers, SciPost Phys. {\bf 7}, 030 (2019).

\bibitem{LL} L. D. Landau, L. P. Pitaevskii, and E. M. Lifshitz, {\it Electrodynamics of Continuous Media, Course of Theoretical Physics Vol. 8} (Pergamon, Oxford, 1984).

\bibitem{Powell-Book} R. C. Powell, {\it Symmetry, Group Theory and the Physical Properties of Crystals} (Springer, New York, 2010).

\bibitem{Database} www.materialsproject.org

\bibitem{Collin-ACB1974} P. Villars and K. Cenzual, {\it Pearson's Handbook of Crystallographic Data for Intermetallic Phases (2nd Edition), Vol. 3} (ASM International, 1996).

\bibitem{Villars2016} P. Villars and K. Cenzual, {\it Rb6[Fe2O5] (Rb6Fe2O5) Crystal Structure: Datasheet from ``PAULING FILE Multinaries Edition -- 2012'' in SpringerMaterials (https://materials.springer.com/isp/crystallographic/docs/ sd{\_}1707360)} (Springer-Verlag Berlin Heidelberg \& Material Phases Data System (MPDS), Switzerland \& National Institute for Materials Science (NIMS), Japan, 2016).

\bibitem{Mumax} A. Vansteenkiste, J. Leliaert, M. Dvornik, M. Helsen, F. Garcia-Sanchez and B. Van Waeyenberge, AIP Advances {\bf 4}, 107133 (2014).

\bibitem{Micromagnum} MicroMagnum - Fast micromagnetic simulator for computations on CPU a and GPU (http://
micromagnum.informatik.uni-hamburg.de). Benjamin Kr\"uger, Ph.D. thesis, Universit\"{a}t Hamburg, 2011.

\bibitem{FN1} For small values of $g$ we did choose smaller sizes of the disk where the magnetization is initially inverted.

\bibitem{FN2} Note that the contribution steeming from the surface density $\vec{n}\cdot\bm{m}$ (here $\vec{n}$ represents the normal to the interfaces of the magnet) can be absorbed, in the case of a thin disk geometry, into the anisotropy constant, $K\rightarrow K_{\textrm{eff}}=K-2\pi M_{s}^{2}$.

\bibitem{Kravchuk2018} V. P. Kravchuk, D. D. Sheka, U. K. R\"{o}ssler, J. van den Brink and Y. Gaididei, Phys. Rev. B {\bf 97}, 064403 (2018).

\bibitem{FN3} In the general case, the  conductivity tensor $\hat{\sigma}$ might also depend on the magnetic configuration due to effects such as anisotropic magnetoresistance.

\bibitem{Thiele} A. A. Thiele, Phys. Rev. Lett. {\bf 30}, 230 (1973).

\bibitem{FN4} The internal force term in the Thiele equation, which reads $F_{i}=\int_{\mathcal{S}} d^{2}\vec{r}\hspace{0.1cm}\bm{H}_{\textrm{eff}}\cdot\partial_{i}\bm{m}$, is identically zero for in-plane skyrmions due to the assumption of translational invariance of the system.

\bibitem{FN5} This value for the damping-like strength of the SOT has been derived from the formula $\tau_{DL}=\gamma\hbar \theta_{SH}/2eM_{s}t$, with $t=1$ nm and $\theta_{SH}=0.15$, see Ref. \onlinecite{Jiang-NatPhys2017}.


\bibitem{Melcher2014} Melcher C, Proc. R. Soc. A {\bf 470}: 20140394 (2014).

\bibitem{Zhang-SciRep2015} X. Zhang, G. P. Zhao, H. Fangohr, J. Ping Liu, W. X. Xia, J. Xia and F. J. Morvan, Sci. Reports. {\bf 5}, 7643 (2015).

\bibitem{Muller-NJP2017} J. M\"{u}ller, New J. Phys. {\bf 19}, 025002 (2017).
\bibitem{Suess-SciRep2019} D. Suess, C. Vogler, F. Bruckner, P. Heistracher, F. Slanovc and C. Abert, Sci. Reports. {\bf 9}, 4827 (2019).

\end{thebibliography}
\end{document}